\documentclass[12pt]{article}

\pdfoutput=1

\usepackage{amsfonts}
\usepackage{epsfig}
\usepackage{latexsym}
\usepackage{graphicx}

\usepackage{epsfig}
\usepackage{latexsym}
\usepackage{graphicx}
\usepackage{amssymb,amsmath}
\usepackage{epstopdf}
\usepackage{epsfig,rotating,amsfonts}

\def\be{\begin{eqnarray}}
\def\ee{\end{eqnarray}}

\def\N{{\cal N}}
\def\pr{\partial}

\def\extd{{\rm d}}

\usepackage{mathptmx}       
\usepackage{helvet}         
\usepackage{courier}        
\usepackage{type1cm}        
\usepackage{makeidx}         

\usepackage{multicol}        
\usepackage[bottom]{footmisc}

\def\spa#1{\phantom{\fbox{\rule[-#1cm]{0cm}{0cm}}}}

\begin{document}

\title{{\Large\bf{A Review of Magnetic Phenomena in Probe-Brane Holographic Matter}\footnote{To appear
in Lect. Notes Phys. ``Strongly interacting matter in magnetic
fields" (Springer), edited by D. Kharzeev, K. Landsteiner, A. Schmitt, H.-U. Yee.}}}
\date{}
\author{Oren Bergman$^a$\footnote{bergman@physics.technion.ac.il}\,,  
Johanna Erdmenger\,$^b$\footnote{jke@mppmu.mpg.de}\, and 
Gilad Lifschytz$^c$\footnote{giladl@research.haifa.ac.il} \spa{0.5}\\
$^a${{\it Department of Physics}},
\\ {{\it Technion - Israel Institute of Technology}},
\\ {{\it Haifa 32000, Israel}}
  \spa{0.5} \\
$^b$ {{\it Max Planck Institute for Physics}}, 
\\ {{\it 80805 Munich, Germany}}
\spa{0.5} \\
$^c$ {{\it Department of Mathematics and Physics}},
\\ {{\it University of Haifa at Oranim}},
\\ {{\it Tivon 36006, Israel}}
}

\maketitle

\begin{abstract}
Gauge/gravity duality is a useful and efficient tool for addressing and studying
questions related to strongly interacting systems described by a gauge theory.
In this manuscript we will review a number of interesting phenomena that occur 
in such systems when a background magnetic field is turned on.
Specifically, we will discuss holographic models for systems that include matter fields
in the fundamental representation of the gauge group, which are incorporated by adding
probe branes into the gravitational background dual to the gauge theory.
We include three models in this review: the D3-D7 and D4-D8 models, that describe four-dimensional
systems, and the D3-D7' model, that describes three-dimensional fermions interacting 
with a four-dimensional gauge field.
\end{abstract}

\newpage

\section{Introduction}

The behavior of strongly interacting matter subject to background magnetic fields is
an interesting and physically relevant problem in many different scenarios,
ranging from the effective 2d electron gas in graphene,
to magnetars, which are neutron stars with a strong magnetic field.
Magnetic fields give rise to a rich array of phenomena.
Some examples in QCD are the magnetic catalysis of 
chiral symmetry breaking \cite{Gusynin:1994re,Gusynin:1994xp,Miransky:2002rp}, anomaly-driven phases 
of baryonic matter \cite{Son:2007ny}, and the chiral magnetic effect
\cite{Fukushima:2008xe}. It has also been suggested that magnetic fields induce
$\rho$-meson condensation and superconductivity in the QCD
vacuum \cite{Chernodub:2010qx,Chernodub:2011mc}. 
There are also many interesting examples in condensed matter physics, most notably the
fractional quantum Hall effect \cite{1982PhRvL..48.1559T,1983PhRvL..50.1395L}.

Gauge/gravity duality, also known as holographic duality, has emerged in recent years
as a particularly useful approach to strong-coupling dynamics.
Although it does not seem to be directly applicable to physical systems, this approach
can be used to study theoretical systems that exhibit the same type of phenomena,
and that capture some of the relevant physics.
The techniques of holographic duality are especially efficient in addressing questions 
associated with finite temperature and density, background fields and transport properties,
that are difficult to study using other non-perturbative methods.
This approach can also lead to new ideas for constructing effective theories of the physical phenomena one is 
interested in. 

Holographic models are divided into two main types, commonly referred to as top-down models and 
bottom-up models. In top-down models the bulk gravitational description of the system
corresponds to a consistent solution of a well-defined quantum gravity theory, either in the context
of the full string theory or in terms of the low-energy effective supergravity theory. 
This then defines some particular strong-coupling boundary dynamics.
In bottom-up models, on the other hand, one builds into the description what one needs
in order to produce the desired boundary dynamics.
Each approach has advantages and disadvantages.
Top-down models are more firmly grounded than bottom-up models,
however they are more restrictive in terms of the variety and scope of phenomena
they can exhibit.

Probe-brane models are a class of top-down holographic models, in which matter fields transforming
in the fundamental representation of a gauge group are incorporated
by embedding ``flavor" D-branes in the gravitational background dual to the gauge theory \cite{Karch:2002sh}. 
These branes are treated as probes, in the sense that we neglect their backreaction on the background.
(This corresponds to the `quenched' approximation in the dual gauge theory, where matter loops are neglected
in computing gluon amplitudes.)
The matter fields are manifest in this construction:
they correspond to the open strings between the flavor branes and the ``color'' branes
that make up the background.
In particular, one can easily design probe brane models in which the light matter degrees of freedom
are purely fermionic, which is obviously a desirable feature for many physical systems,
including QCD and condensed matter electron systems. 

The matter dynamics is determined by the properties of the probe brane embedding.
In particular, the fluctuations of the probe brane worldvolume fields correspond to gauge-invariant
composite operators that describe the mesonic states of the matter system.
There are generically two types of embeddings at finite temperature:
``BH embeddings", in which the brane extends to the horizon of the background,
and ``MN embeddings", in which the brane terminates outside the horizon.
The two embedding types describe different phases of the matter in the dual gauge theory.
For example, in the MN phase the mesons are stable
since they are associated with real eigenfrequencies of the probe
brane fluctuations. In the BH phase, on the other hand, some of the energy of the fluctuations
is dissipated into the black hole, leading to complex eigenfrequencies
and damping. In this case the mesons have a finite lifetime. 
MN embeddings are favored at low temperature, and 
as the temperature is increased one generically observes
a first order phase transition to a BH embedding.

The most extensively studied probe-brane models are the D3-D7 model 
\cite{Karch:2002sh},
in which D7-branes are added to the D3-brane background, 
and the D4-D8 (or Sakai-Sugimoto) model \cite{Sakai:2004cn}, in which D8-branes and anti-D8-branes 
are added to the background of D4-branes compactified on a circle.
Both models describe a strongly-coupled gauge theory in four dimensions with fundamental matter degrees of freedom,
and both exhibit a number of phenomena similar to QCD.
More recently, a different D3-D7 system, more closely related to the D4-D8 system, has been
used as a model of strongly-interacting fermionic matter in three spacetime dimensions
\cite{Rey_strings,Rey:2008zz,Davis:2008nv,Bergman:2010gm}.
We will refer to this as the D3-D7' model.
This model exhibits several interesting phenomena that are familiar in planar condensed matter systems.

Probe-brane models are especially well-designed to study the properties of the dual matter systems
at non-zero density and in background electromagnetic fields.
Both are implemented by turning on specific components of the probe-brane worldvolume gauge
field, and solving the resulting coupled differential equations for the embedding and the gauge fields.
Here one observes another basic difference between the two types of embeddings in terms
of their response to a background electric field.
MN embeddings correspond to electrical insulators with a mass-gap to charged excitations,
and BH embeddings describe gapless conductors.

Probe-brane models also exhibit a number of interesting phenomena in a background magnetic field,
which are qualitatively similar to the phenomena listed in the beginning.
In this paper we will review how each of the three models mentioned above respond at non-zero
density to a background magnetic field in various situations.
In particular, we will encounter the magnetic catalysis effect in both the D3-D7 and D4-D8 models.
In the D3-D7 model we will also demonstrate the formation of a superfluid state. 
In the D4-D8 model we will describe anomaly-generated currents and baryonic states,
as well as a metamagnetic-like transition. 
In the D3-D7' model we will see both quantum and anomalous Hall effects, as well
as how the magnetic field influences the instability to the formation of stripes,
and the zero-sound mode.

This paper is divided into three main sections, reviewing each of 
the probe-brane models in turn. For completeness let us mention that
magnetic fields also play an important role in
the related D3-D5 system, where the
probe D5-brane corresponds to additional $(2+1)$-dimensional degrees
of freedom in the dual gauge theory. In this model,
the magnetic field leads to a phase transition of
Berezinskii-Kosterlitz-Thouless (BKT) type
\cite{Jensen:2010ga,Evans:2010hi}. For brevity we do not discuss this
model in this review.

\section{The D3-D7 model}

\subsection{Brane construction}

The starting point for this model is the usual configuration of the
AdS/CFT correspondence \cite{Maldacena:1997re} which involves a stack of $N$ D3
branes. This has an open string interpretation in which the low-energy
degrees of freedom are described by $U(N)$ $\N=4$ Super-Yang-Mills
theory. On the other hand, in the closed string interpretation of $N$
D3 branes, the low-energy {\it near-horizon} limit gives rise to the
space $AdS_5 \times S^5$. Identifying the two pictures leads to the
AdS/CFT correspondence.

Let us now add $N_f$ probe D7-branes to this configuration, 
as first done in \cite{Karch:2002sh} and reviewed in detail in 
\cite{Erdmenger:2007cm}. Within 
(9+1)-dimensional flat space, the D3-branes  are extended along the
0123 directions,
whereas the D7-branes are extended along the  01234567 directions.
This configuration preserves $1/4$ of the total amount of
supersymmetry in type IIB string theory (corresponding to 8 real
supercharges) and has an $SO(4) \times SO(2)$ isometry in the
directions transverse to the D3-branes. The $SO(4)$ rotates $x^4, x^5,
x^6, x^7$, while the $SO(2)$ group acts on $x^8, x^9$. Separating the
D3-branes from the D7-branes in the (8,9) directions by a distance $l$
explicitly breaks the $SO(2)$ group. 
These geometrical symmetries are also present in the dual field theory:
The dual field theory is an $\N=2$ supersymmetric 
(3+1)-dimensional theory in which the degrees of freedom of 
$\mathcal{N}=4$ super
Yang-Mills theory are coupled to $N_f$ hypermultiplets of 
flavor fields with fermions and scalars 
$(\psi_i, q^n)$, $i=1,2$, $n=1, 2 $, which
transform in the fundamental representation of the gauge group.
Separating the D7-branes from the D3-branes corresponds to giving a mass
to the hypermultiplets.

For massless flavor fields, the Lagrangian is
classically invariant under conformal transformations
$SO(4,2).$\footnote{However note that the scale-invariance is broken
  at the quantum level since the beta function is proportional to $N_f
  / N_c$ and therefore non-vanishing. In the limit $N_c \rightarrow
  \infty$ with $N_f$ being fixed, 
  the beta function is approximately zero, i.e.~we may treat the theory
  as being scale invariant also at the quantum level.}  Moreover, 
the theory is invariant under 
 the ${R}$-symmetries $SU(2)_{R}$ and
 $U(1)_{R}$ as well as  under the global $SU(2)_{\Phi}$, which rotates the scalars in the adjoint hypermultiplet. 
 Note that the mass term in the Lagrangian breaks the $U(1)_{{R}}$ symmetry explicitly. 
If all $N_f$ flavor fields have the same mass $m,$ the field theory is invariant under a global $U(N_f)$ flavor group. 
The baryonic $U(1)_B$ symmetry is a subgroup of the $U(N_f)$ flavor group. 
These symmetries of the field theory side may be identified with
 symmetries of the D3-D7 brane intersection and hence also with the dual
 gravity description. 

For this field theory, gauge invariant composite operators may now be
constructed which transform in suitable representations of the $SU(2)
\times SU(2) \times U(1)$ symmetry group isomorphic to the
geometrical
$SO(4) \times SO(2)$.  These operators are  expected to
be dual to the fluctuations of the D7-brane which transform in the same
representation, as worked out in detail in 
\cite{Kruczenski:2003be}. An example of a  meson operator is given by
\begin{equation} \label{eq:M}
{\cal M}^A = \bar \psi_i \sigma^A{}_{ij} \psi_j + \bar q^m X^A q^m \, ,\qquad (i,m=1,2) ,
\end{equation}
with $X^A$ the vector $(X^8,X^9)$ of  adjoint scalars associated with the $(8,9)$ directions,
and $\sigma^A \equiv (\sigma^1,\sigma^2) $ a doublet of Pauli
matrices. Thus \eqref{eq:M}
 has charge $+2$ under $U(1)_R$. It is a singlet under both $SU(2)_\Phi$ and $SU(2)_R$.
The conformal dimension is $\Delta=3$. 
This operator may be viewed as a supersymmetric generalization of a
mesonic operator in QCD, with the index $A$ labelling two scalar mesons.

The standard AdS/CFT duality relates the $\N=4$ Super Yang-Mills degrees of freedom to
supergravity on $AdS_5\times S^5$.
In addition, there are new degrees of freedom associated to the D7-brane
worldvolume fields
originating from the open strings on the D7-brane. The additional duality maps
these to the
mesonic operators in the field theory. This is an open-open string duality, as
opposed to the
standard AdS/CFT correspondence, which is an open-closed string duality.
 The dynamics of the D7-brane is described by the Dirac-Born-Infeld (DBI) action
\begin{equation}
S_{D7} = -\frac{\mu_7}{g_s}\int \mathrm{d}^8\xi\,
\sqrt{-\det\left(G_{ab} + B^{(2)}_{ab}+ 2\pi\alpha' F_{ab}\right)} \,,
\label{daction}
\end{equation}
where $\mu_7=[(2\pi)^7\alpha'^4]^{-1}$. 
$G$ and  $B^{(2)}$ are the induced metric and  two-form
field on the probe brane worldvolume, and
$F_{ab}$ is the worldvolume field strength. 
The D7-brane action also
contains a fermionic term $S^f_{D7}$. In addition there may
also be contributions of Wess-Zumino form. An example for this will be
discussed below.

Let us write the $AdS_5 \times S^5$ metric in the form
\begin{equation} \label{AdSmetric}
ds^2 \, = \, \frac{r^2}{R^2} \eta_{ij} dx^i dx^j
 \, + \, \frac{R^2}{r^2} (d \rho^2 + \rho^2 d \Omega_3^2 + dx_8^2 + dx_9^2)
 \, ,
\end{equation}
with $i,j=0,1,2,3$, $\rho^2=x_4^2+...+x_7^2$, $r^2=\rho^2+x_8^2+x_9^2$
and $R$ the AdS radius. Since the D7-brane is transverse to $x_8$,
$x_9$ in flat space, we see that it extends
along $AdS_5$
and wraps an $S^3$ inside $S^5$ in the near-horizon background. The
action for a static D7-brane embedding, for which $F_{ab}$ may be consistently set to zero on its
world-volume, is given from (\ref{daction})  up to angular factors by
\begin{equation}
S_{D7} = - \frac{\mu_7}{g_s}\int \mathrm{d}^8\xi\, \rho^3
\sqrt{1 + \dot{x}_8^2 + \dot{x}_9^2} \,,
\end{equation}
where a dot indicates a $\rho$ derivative ({\em e.g.}\ $\dot x_8 \equiv
\partial_\rho x_8$). The ground state configuration of the D7-brane then
corresponds to the solution of the equation of motion
\begin{equation} \label{d7eom} {\frac{d}
{d \rho}} \left[ {\rho^3 \over \sqrt{1 + {\dot x_{8}}^2+ {\dot x_{9}}^2}
}{ d x \over d \rho}\right] = 0 \, ,
\end{equation}
where $x$ denotes either $x_8$ or $x_9$. Clearly the action is
minimized by $x_8,x_9$ being any arbitrary constant. Therefore the embedded D7-brane is flat.
According to string theory, the choice of the position in
the $x_8,x_9$ plane corresponds to choosing the quark mass in the
gauge theory action.  The fact that $x_8,x_9$ are constant at all values of
the radial coordinate $\rho$, which corresponds to the holographic renormalization scale, may be interpreted as non-renormalization of the mass in the dual field theory. 

In general, the equations of motion have asymptotic ($\rho\rightarrow \infty$)
solutions of the form
\begin{equation} \label{probeasymptotic}
x = l + \frac{c}{\rho^2} + ... \,,
\end{equation}
where $l$ is related to the quark mass $m$ by
\begin{gather}   m= \frac{l}{2 \pi \alpha' } \, .\end{gather}
 In agreement with the standard AdS/CFT
result about the asymptotic behaviour of supergravity fields near the boundary, the parameter
$c$ must correspond to the vev of an operator with the same symmetries as the
mass and of dimension three, since $\rho$ carries energy dimension.
$c$ is therefore a measure of the quark condensate $\tilde{\psi} \psi$.
$c$ is obtained from $\partial {\cal L} / \partial m$ which in addition to the fermion bilinear
also includes scalar squark terms.
We  may consistently assume that the squarks have zero vev. 
Moreover, supersymmetry requires that a vev for $c$ must be absent since $c$
is an F-term of a
chiral superfield:  $
  \tilde  \psi \psi$ is the F-term of $ \tilde   Q  Q$.
  Supersymmetry is broken if $c=\langle \tilde  \psi  \psi \rangle \neq
  0$. This is reflected also in the supergravity solution:
The solutions to the supergravity equations of motion with $c$ non-zero are not regular in AdS space
and  are therefore excluded. 

We therefore consider the regular supersymmetric embeddings of the D7-brane
for which the quark mass $m$ may be non-zero, but the condensate $c$ vanishes. 
For massive embeddings, the D7-brane is
separated from the stack of D3-branes in either the $x_8$ or $x_9$
directions, where the indices refer to the coordinates given in
\eqref{AdSmetric}. In this case the radius of the $S^3$
becomes a function of the radial coordinate $r$ in ${\rm AdS}_5$. At a
radial distance from the deep interior of the AdS space given by the
hypermultiplet mass, the radius of the $S^3$ shrinks to zero. From a
five-dimensional AdS point of view, this gives a minimal value for the radial coordinate $r$ 
beyond which the D7-brane cannot extend further. 
This is in agreement with 
the induced metric on the D7-brane world-volume, which is given by
\begin{align} \label{inducedmetric}
ds^2 \, & = \, \frac{\rho^2+l^2}{R^2} \eta_{ij} dx^i dx^j
 \, + \, \frac{R^2}{\rho^2 + l^2} d\rho^2
 \, + \, \frac{R^2 \rho^2}{\rho^2+l^2 }  d\Omega_3{}^2\, , \\ 
d \Omega_3{}^2 & = d \psi^2 + \cos^2 \psi d\beta^2 + \sin^2 \psi d
\gamma^2 \, , 
\end{align}
where $\rho^2 = r^2 -l^2$ and $\Omega_3$ are spherical coordinates in
the 4567-space. For $\rho \rightarrow \infty$,
this is the metric of  $AdS_5 \times S^3$.  When
$\rho=0$ ({\em i.e.}~$r^2=l^2$), the radius of the $S^3$ shrinks to zero.

\subsection{Finite temperature}

The finite temperature system is realized holographically by placing the D7-brane
in an AdS-Schwarzschild black hole background with metric given by
\begin{eqnarray}
\label{eq:AdSS}
ds^2 &=& \frac{r^2 }{2R^2} \left(\frac{\extd \vec{x}^2
   \left(r_h{}^4+r ^4\right)}{r^4}-\frac{\text{dt}^2
   \left(r^4-r_h{}^4\right)^2}{r^4 \left(r_h{}^4+r^4\right)}\right) \nonumber \\  
 && \qquad \mbox{} + \frac{R^2}{r^2}\left(\text{d}L^2+\text{d$\rho $}^2+ L^2 \text{d$\phi $}^2
+\rho ^2 \text{d$\Omega $}_3^2\right)\, ,
\end{eqnarray}
where $r^2=\rho^2+L^2$. 
The temperature is given by $T=r_h / {(\pi R^2)} $.

In this metric we have introduced polar coordinates $(L,\phi)$ in the
$(x_8,x_9)$ plane and consider solutions for D7-brane embeddings with
$L=L(\rho)$, $\phi=\mathrm{const}$. The asymptotic near-boundary
behaviour of these brane embeddings is given by
\begin{gather}  \label{Lrho}
L (\rho) = l + \frac{c}{\rho^2} + \dots \,  ,
\end{gather}
where $m= l/(2 \pi \alpha')$ is the bare quark mass and $c$ is proportional to
the condensate $\langle \bar \psi \psi  \rangle$.
At finite temperature, supersymmetry is broken and brane embeddings
with $c$ non-zero are possible, in
contrast to the supersymmetric case discussed above.

Depending on the quark mass, there are two different types of
embeddings: Those which reach the black hole, and those which do not
since the $S^3$ they wrap shrinks to zero outside the black hole horizon.
The first type of branes is referred to as `black hole' (BH)
embeddings, while the second type is referred to as `Minkowski' (MN) embeddings.
In the BH case,
fluctuations of the probe brane have complex eigenfrequencies or
quasi-normal modes, which
means that the mesons associated with these fluctuations decay. In the
MN phase, the mesons are stable. The phase transition between the
two types of embeddings is first order.

At finite temperature, the solution with $m=0$ also has
$c=0$. However, in gravity backgrounds corresponding to confining
field theories, brane embeddings with $m=0$, $c \neq 0$ are
possible. These realize spontaneous chiral symmetry breaking
\cite{Babington:2003vm}. 

\subsection{Magnetic catalysis}

Let us now consider, as was first done in \cite{Filev:2007gb}, a magnetic
field induced by a pure gauge B-field in the worldvolume direction of
the D3-branes,
\begin{gather} \label{2form}
B^{(2)} = B dx^2 \wedge dx^3 \, ,
\end{gather}
which satisfies $dB^{(2)}=0$. This contributes to the DBI action
\eqref{daction}. Since $B^{(2)}$ and $2 \pi \alpha' F$ enter
\eqref{daction} in the same way, we may trade $B^{(2)}$ for a gauge
field on the probe brane via $F = -B^{(2)}/2 \pi \alpha'$, 
which justifies interpreting $B$
of \eqref{2form} as a magnetic field.

In addition, there is a non-trivial Wess-Zumino
contribution to the action, which at first order in $\alpha'$ is of the form
\begin{gather}
S_\mathrm{WZ} = 2 \pi \alpha' \mu_7 \int \! F \wedge C^{(6)} \,.
\end{gather}
In the presence of the B-field, this leads to an additional non-trivial
contribution to the action, as explained in detail in
\cite{Filev:2007gb}. This gives rise to a non-trivial $C^{(6)}$ which
breaks supersymmetry on the worldvolume of the D7-brane.

\begin{figure}[ht]
\begin{center}
\includegraphics[width=8cm,clip=true,keepaspectratio=true]{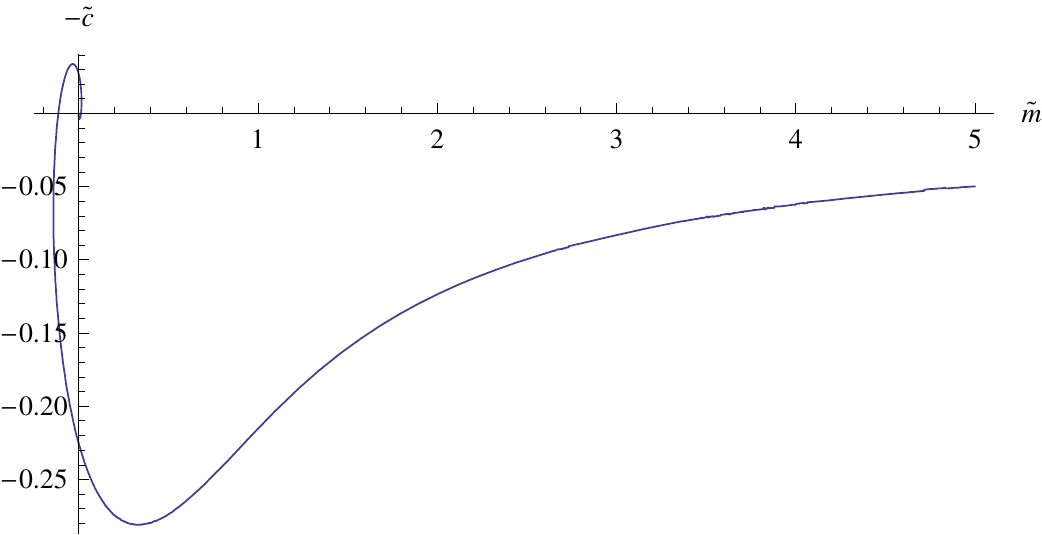}
\caption{$-\tilde{c}(\tilde{m})$ at vanishing temperature, 
with $- \tilde c= \langle \bar \psi \psi \rangle (2 \pi \alpha')^3 /
(R^3 B^{\textstyle \frac{3}{2}})$, $\tilde m = m (2 \pi \alpha')/(R \sqrt{B})$.
  Reproduced from \cite{EMS}.}
\label{fig:spiral}
\end{center}
\end{figure}

Since supersymmetry is broken, the D7-brane now has a profile which
depends on $\rho$ as in \eqref{Lrho}, even at zero temperature. 
The Lagrangian corresponding to \eqref{daction} takes the form
\begin{gather} {\cal L} = 
- \frac{\mu_7}{g_s} \rho^3 \sin \psi \cos \psi \sqrt{1 + L'^2}
\sqrt{
  1+ \frac{R^4 B^2}{(\rho^2 + L^2)^2}} \, .
\end{gather}
For $m=0$, the brane embedding solution obtained from this Lagrangian has non-zero $c \propto
\langle \bar \psi \psi \rangle$ in \eqref{Lrho}.
The magnetic field therefore induces spontaneous chiral
symmetry breaking, a phenomenon known as {\it magnetic catalysis} \cite{Gusynin:1994re,Gusynin:1994xp,Miransky:2002rp}. 
For large $m$, the condensate may be calculated analytically
\cite{Filev:2007gb} and is found to be
\begin{gather}
\langle \bar \psi \psi  \rangle \propto  - c = - \frac{R^4}{4l} B^2 \, .
\end{gather}
For small $m$, $c$ has to be evaluated numerically. The result is
shown in figure \ref{fig:spiral}. By evaluating the free energy, it
has been shown that when there is more than one solution, the one
with the larger condensate is preferred. 
  
At small values of the magnetic field, it is possible to analytically
evaluate the shift of the meson masses due to its presence. For
fluctuations of the embedding scalar in particular, a Zeeman splitting
is observed \cite{Filev:2007gb}. While in the absence of a magnetic field, in
the supersymmetric case described here, the scalar meson mass obtained from
the fluctuations is $M_0(n)=  2m/\sqrt{\lambda} \cdot \sqrt{(n+1)(n+2)}$ \cite{Kruczenski:2003be}, for non-zero
magnetic field there is a mass splitting
\begin{gather}
M_\pm = M_0 \pm \frac{1}{\sqrt{\lambda}} \frac{B}{m} \, .
\end{gather}

A review of magnetic catalysis in probe D7-brane systems is given in 
\cite{Filev:2010pm}. Magnetic catalysis is also found in systems involving $N_f$ D7 branes
where the backreaction of the metric on the background geometry is
taken into account \cite{Filev:2011mt,Erdmenger:2011bw,Ammon:2012qs}. 
Out-of-equilibrium dynamics associated with the phase transition
induced by magnetic catalysis has been investigated in \cite{Evans:2010xs}.

\begin{figure}[ht]
\begin{center}
\includegraphics[width=12cm,clip=true,keepaspectratio=true]{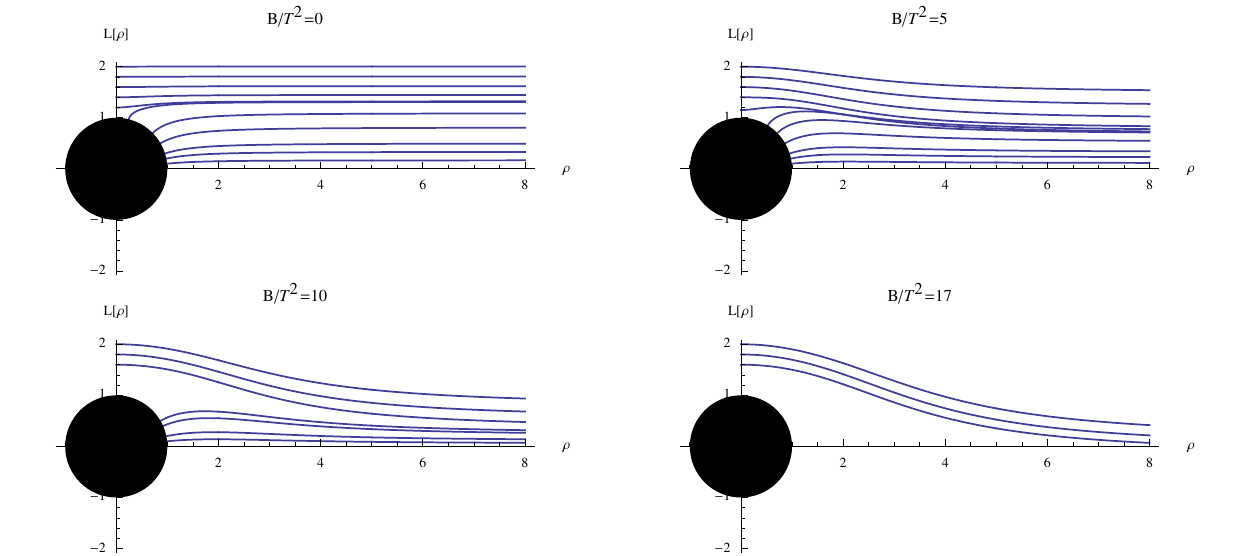}
\caption{Increasing values of ${B}/T^2  $ for
fixed $T$ show the repulsive nature of the magnetic  field.
We see that for
large enough ${B}/T^2$, the melted 
phase is never reached, and the chiral symmetry is spontaneously
broken. Figure reproduced from \cite{EMS}.}
\label{fig:BandTflowsv}
\end{center}
\end{figure}

In the finite temperature case, there is a
competition between two mechanisms: The black hole attracts the D7-brane, 
while it is repelled at small radii by the magnetic field. This
implies a phase transition between a phase where the D7-brane reaches
the black hole and one where it does not. 
This is shown in figure \ref{fig:BandTflowsv}
for different values of the magnetic field, where the dimensionless
ratio $B/T^2$ is used. A detailed discussion of the normalization is
found in \cite{EMS}.

\begin{figure}[ht]
\begin{center}
\includegraphics[width=8cm]{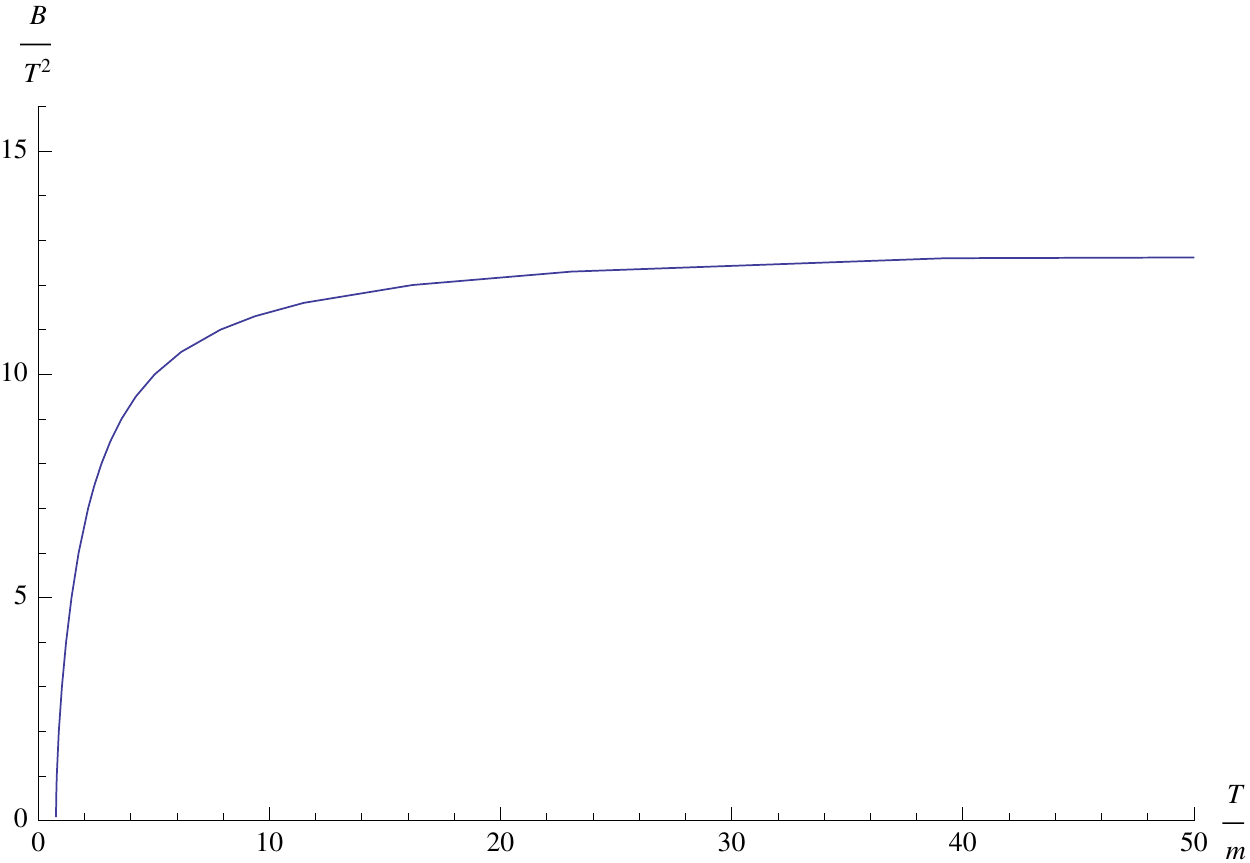}
\caption{Phase diagram for the D3/D7 system in the
$(B/T^2,T/m)$ plane. Figure from \cite{EMS}.}
\label{fig:phasediagBT} 
\end{center}
\end{figure}

There is a critical value for $B/T^2$  above which the probe brane is repelled from the black hole for
all values of the bare quark mass: In this case, chiral symmetry is
spontaneously broken and the mesons are stable. The phase diagram is
shown in figure \ref{fig:phasediagBT}. 

\begin{figure}[ht]
\begin{center}
\includegraphics[width=9cm,clip=true,keepaspectratio=true]{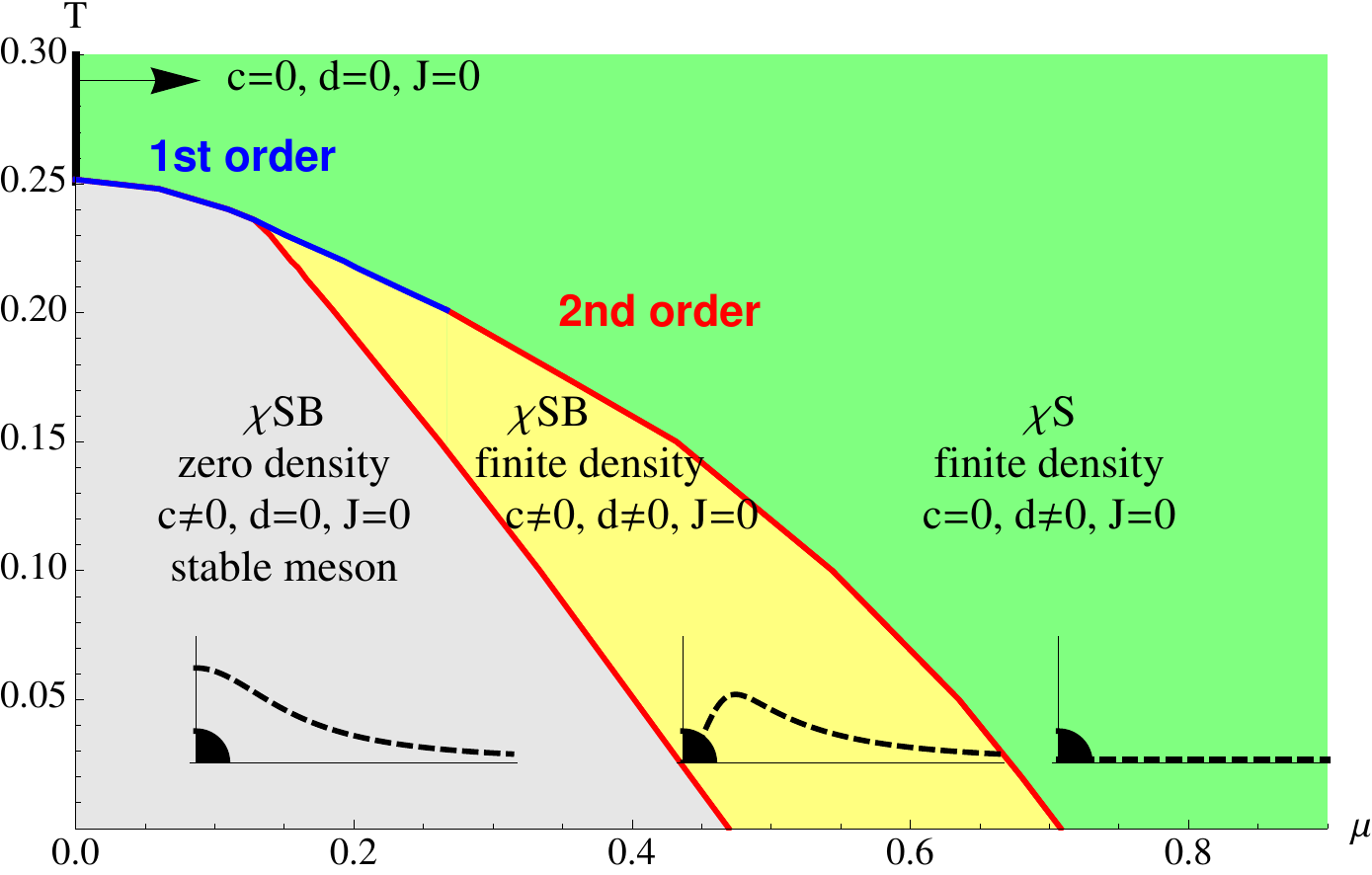}
\caption{Phase transitions at constant $B$ field in the $(T,\mu)$
  plane for the D3/D7 model. Figure reproduced from
  \cite{Evans:2011mu} by kind permission of the authors. $c$, $d$, $J$
refer to the condensate, density and electric current, respectively.}
\label{fig:Bmu}
\end{center}
\end{figure}

More involved phase diagrams are obtained if a $U(1)$ chemical potential and density
are turned on in addition to the magnetic field by considering a
non-trivial profile for the $U(1)$ gauge field on the D7-brane
\cite{Evans:2010iy,Jensen:2010vd,Evans:2011mu}. An example of a phase diagram is
shown in figure \ref{fig:Bmu}. 

\subsection{Superfluid}

At finite isospin
density, the D3-D7 model realizes a holographic superfluid \cite{Ammon:2008fc,Ammon:2009fe}. 
Finite isospin
density is obtained by considering two coincident D7-branes, and
using an ansatz for solving the equations of motion,
which involves a non-trivial profile for the the temporal component of
the $SU(2)$ worldvolume gauge field, with asymptotic behavior 
\begin{gather} \label{A3rho}
A^3{}_t (\rho) \propto \mu^3
 + \frac{d^3}{\rho^2} \, .
\end{gather}
$\mu^3$ breaks the $SU(2)$ symmetry explicitly to a residual $U(1)_3$. 
In the presence of this background, 
the energetically favored solution also involves a non-trivial spatial
component of the worldvolume gauge field,
\begin{gather}
A^1{}_x (\rho) \propto \frac{d^1{}_x}{\rho^2} \, .
\end{gather}
Here the leading contribution is absent in the asymptotic behavior,
so the $U(1)_3$ symmetry is spontaneously  broken.
$A^1{}_x$ is dual to a condensate of the form
\begin{gather}
d^1{}_x \propto \langle  \bar \psi \sigma^1 \gamma_x \psi +
\bar \phi \sigma^i \pr_x \phi \rangle \,  ,
\end{gather}
which is the supersymmetric equivalent of the $\rho$ meson.
The calculation of the frequency-dependent conductivity
$\sigma(\omega)$ for this
solution shows that it describes a superfluid: $\sigma(\omega)$
displays a gap. For the Sakai-Sugimoto model discussed below, a
similar condensation mechanism has been found in \cite{Aharony:2007uu}
and superfluidity has been discussed in \cite{Rebhan:2008ur}.

\begin{figure}[ht]
\begin{center}
\includegraphics[width=7cm,clip=true,keepaspectratio=true]{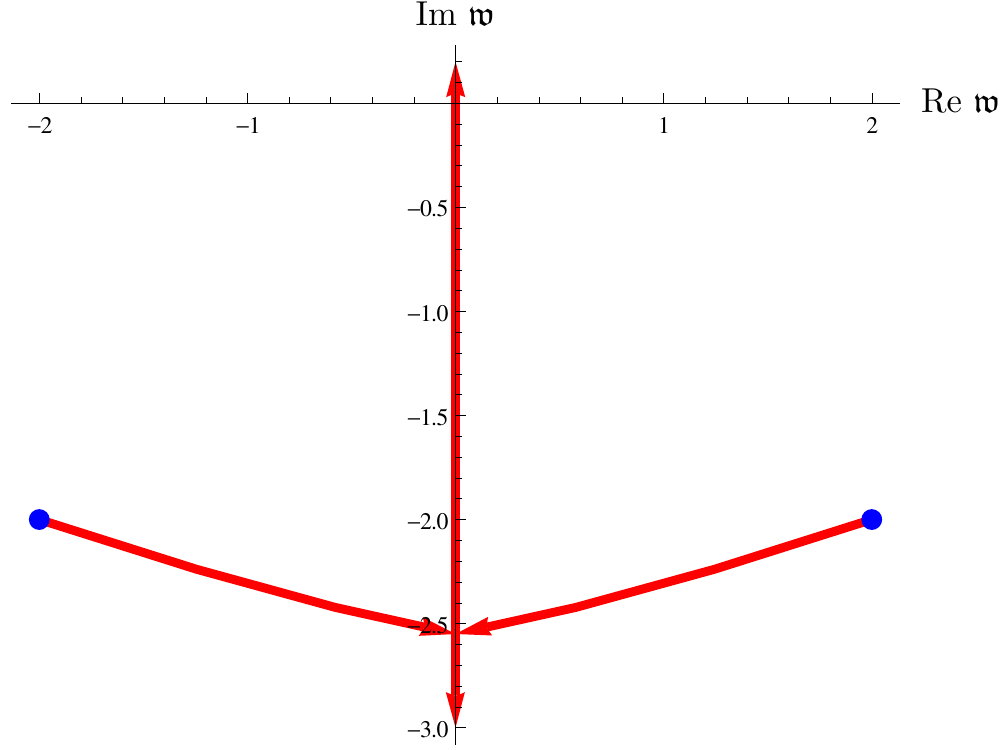}
\caption{Quasi-normal modes cross into the upper half plane above a
  critical magnetic field, signalling an instability.  Figure
  reproduced from \cite{Ammon:2011je}.}
\label{fig:Bcrit}
\end{center}
\end{figure}

As discussed in \cite{Erdmenger:2010zz,Ammon:2011je}, a similar condensation process also happens
when the profile \eqref{A3rho} for the {\it temporal} component of the
$SU(2)$ gauge
field is replaced by a non-trivial profile for a {\it spatial}
component of the form
\begin{gather}
A^3{}_x = B y \, ,
\end{gather}
which corresponds to a background magnetic field. In this case a similar
condensation as above takes place. This has been demonstrated by
analyzing the fluctuations 
about the magnetic field background \cite{Ammon:2011je}: The
quasi-normal modes of particular fluctuations cross into the upper
half of the complex frequency plane above a critical value of the
magnetic field, indicating an instability. This is shown in figure
\ref{fig:Bcrit}. Finally let us note that the Hall conductivity has
been calculated for the D3/D7 model in \cite{O'Bannon:2007in}. 
Unlike the isospin case, the ground
state involving the $\rho$ condensate is spatially modulated for the
magnetic field background, leading
to an Abrikosov lattice \cite{new}. A similar $\rho$ meson condensation mechanism
in a background magnetic field has been
found in the context of field theory in \cite{Chernodub:2010qx,Chernodub:2011mc,Braguta:2011hq}, based on
similar earlier results in electroweak theory \cite{Ambjorn:1989sz}. For the Sakai-Sugimoto model which we
discuss below, a similar mechanism has been discussed in
\cite{Callebaut:2011uc,Callebaut:2011ab}.

\section{The D4-D8 (Sakai-Sugimoto) model}
\label{D4D8_section}

\subsection{Basics}

$N_c$ D4-branes on $\mathbb{R}^{1,3}\times S^1$ with anti-periodic boundary conditions for fermions provide 
a holographic model for the low energy behavior of 4d $SU(N_c)$ Yang-Mills theory with 
$g_{YM}^2 = 4\pi g_s\sqrt{\alpha'}/R_4$ \cite{Witten:1998zw}.
The near-horizon background at zero temperature
is given by (we work with dimensionless coordinates rescaled by $R$)
\be
\label{confined_background}
ds^2_{con} &=& u^{3\over 2} \left(-dx_0^2 + d{\bf x}^2
+ f(u) dx_4^2\right)
+ u^{-{3\over 2}}\left({du^2\over f(u)}
+ u^2 d\Omega_4^2\right) \nonumber\\
e^{\Phi} &=& g_s u^{3/4} \; , \;
F_4 = 3\pi (\alpha')^{3/2}  N_c \, d\Omega_4 \,,
\ee
where $f(u) = 1 - (u^3_{KK}/u^3)$, $u_{KK} = 4R^2/(9R_4^2)$ and $R = (\pi g_s N_c)^{1/3} \sqrt{\alpha'} $.
The IR ``wall" at $u=u_{KK}$ implies that the dual gauge theory is confining.
At nonzero temperature there is another possible background with a metric
\be
\label{deconfined_background}
ds^2_{dec} = u^{3\over 2} \left(-f(u)dx_0^2 + d{\bf x}^2
+ dx_4^2\right)
+ u^{-{3\over 2}}\left({du^2\over f(u)} 
+ u^2 d\Omega_4^2\right)\,,
\ee
where $f(u) = 1 - (u^3_{T}/u^3)$ and $u_T=(4\pi/3)^2 R^2 T^2$.
This background becomes the dominant one when $T>1/(2\pi R_4)$.
The presence of a horizon at $u=u_T$ in this background indicates that the gauge theory
undergoes a (first order) deconfinement transition at this temperature.

Quarks are added to the model by including D8-branes and anti-D8-branes that are localized on the circle 
\cite{Sakai:2004cn}.
With $N_f$ D8-branes at one point and $N_f$ anti-D8-branes at another point, the model has $N_f$ flavors
of massless right-handed and left-handed fermions, and a $U(N_f)_R\times U(N_f)_L$ chiral symmetry.
The 8-branes are treated as probes in the near horizon background of the D4-branes. 
The flavor dynamics is thus encoded in the 5d effective worldvolume theory of the 
D8-branes, which includes a DBI term and a CS term 
(in Lorentzian signature)\footnote{For simplicity, we will consider the 
single flavor case with one D8-brane and one anti-D8-brane.
This does not affect any of the results qualitatively.}
\be
\label{DBI_action}
S_{DBI} &=&  - {\cal N} \int d^{4}x\, du \, u^{1/4} \sqrt{-\mbox{det}(g_{MN} + f_{MN})} \\
\label{CS_action}
S_{CS} &=& - \frac{\cal N}{8} \int d^4x\, du \, \epsilon^{MNPQR} a_M f_{NP} f_{QR} \,.
\ee
The dimensionless worldvolume gauge field $a_M$ 
and field strength $f_{MN}$ are defined as
$a_M = \linebreak[4] (2\pi\alpha'/R) A_M$ and  $f_{MN} = 2\pi\alpha' F_{MN}$, and the overall
normalization is given by 
${\cal N} = \mu_8 \Omega_4 R^9/g_s = (1/3) N_c\, R^6 (2\pi)^{-5}(\alpha')^{-3}$.
The anti-D8-brane has a similar action in terms of its worldvolume gauge field $\bar{a}_M$.
The DBI term is identical to that of the D8-brane, and the CS term has the opposite sign.
We define the vector combination as $a^V_M = \frac{1}{2}(a_M + \bar{a}_M)$,
and the axial combination as $a^A_M = \frac{1}{2}(\bar{a}_M - {a}_M)$.

In the low-temperature confining background (\ref{confined_background})
the D8-brane and anti-D8-brane connect at $u=u_0\geq u_{KK}$
into a smooth U-shaped configuration (Fig.~\ref{embeddings}a), reflecting the spontaneous breaking
of the $U(1)_R\times U(1)_L$ chiral symmetry to the diagonal $U(1)_V$.
The embedding is determined by the DBI action (setting $f_{MN}=0$)
\be
S_{DBI}^{con} =  - {\cal N} \int d^4x\, du \, u^4 \left[f(u)(x_4'(u))^2 + {1\over u^3f(u)}\right]^{\frac{1}{2}} \,,
\ee
which implies an asymptotic behavior
\be
\label{U_embedding}
x_4(u) \approx {L\over 2} - {2\over 9} {u_0^4\sqrt{f(u_0)}\over u^{9/2}} \,,
\ee
where $L$ is the asymptotic brane-antibrane separation. 

The normalizable fluctuations of the D8-brane worldvolume fields in this embedding correspond 
to the (low spin) mesons of the model.
Their mass scale is set by $u_0$, which we can think of as the mass of a ``constituent quark" described by 
an open open string from $u_0$ to $u_{KK}$.
There is one massless pseudoscalar field $\varphi$, 
precisely as one expects from the broken chiral symmetry.
This is related to the $\eta'$ meson in QCD.
It appears (in a gauge with $a_u=0$) as the zero mode $a^A_\mu$:\footnote{Note that, although 
the boundary value is non-zero, this is a normalizable mode since the field strength is normalizable.
Ordinarily, boundary values of bulk fields correspond to parameters in the boundary theory.
But in this case there is a possible ambiguity, since the boundary value of $a^A_\mu$ can also describe
a non-trivial gradient of the pseudoscalar field.}
\be
\label{pseudoscalar}
a_\mu^A(x^\mu,u) = - \partial_\mu \varphi(x^\mu) \psi_0(u) + 
\mbox{higher modes} \,,
\ee
where 
\be
\psi_0(u) = {2\over\pi}\, \arctan\sqrt{{u^3\over u_{KK}^3}-1}\,.
\ee

Baryons are described by D4-branes wrapped on $S^4$ inside the D8-brane.
Their charge comes from the $N_c$ strings which must be attached to the wrapped D4-brane to cancel 
a tadpole due to the background RR field. These strings end on the D8-brane, giving $N_c$ units of 
charge.\footnote{For $N_f>1$ the baryons correspond to instantons in the non-abelian D8-brane theory 
\cite{Sakai:2004cn,Hata:2007mb}.
This reproduces the known description of baryons as Skyrmions in the chiral Lagrangian.
In this description the baryon charge comes from the CS term coupling the $U(1)_V$ field to the 
instanton density in the $SU(N_f)_V$ part.} 

In the high-temperature deconfining background (\ref{deconfined_background}) 
the D8-branes and anti-D8-branes can be either connected (Fig.~\ref{embeddings}b) or 
disconnected, with $x_4(u)=L/2$ (Fig.~\ref{embeddings}c),
the latter corresponding to the restoration of the chiral symmetry \cite{Aharony:2006da}.
The DBI action in the high-temperature deconfining background is very similar: 
\be 
\label{D8_action_deconfined}
S_{DBI}^{dec} =   - {\cal N} \int d^4 x\, du \, u^4 
\left[ f(u) (x_4^\prime(u))^2 
 + \frac{1}{u^3} 
\right]^{1\over 2} \,.
\ee
Consequently the properties of the U embedding in this background are qualitatively similar 
to those of the embedding in the confining background, for example in terms of the spectrum of mesons.
In the disconnected embedding there are no normalizable fluctuations corresponding to mesons, as one expects 
in a chiral-symmetric phase.
Comparing the (Euclidean) actions of the two embeddings shows that the disconnected one
becomes dominant when $T>0.154/L$.
In particular, for small $L$ ($L<0.97 R_4$) the gauge theory has an intermediate phase 
of deconfinement with broken chiral symmetry.

\subsection{Finite density and background fields}

The D8-brane worldvolume vector and axial gauge fields are dual to conserved vector and axial 
currents in the gauge theory, and therefore\footnote{The axial symmetry is broken by an anomaly.
However this is a subleading effect at large $N_c$ which we can neglect.
In particular, we will assume that the one-flavor pseudoscalar $\eta'$ is massless.
For a discussion of the $U(1)_A$ anomaly and the $\eta'$ mass in the context of the Sakai-Sugimoto model
see \cite{Sakai:2004cn,Bergman:2006xn}.}
\be
\label{conserved_currents}
j^\mu_{V,A} =  \frac{1}{{\cal N}V_4} \frac{\partial S_{D8}|_{on-shell}}{\partial a_\mu^{V,A}(u\rightarrow\infty)} \,.
\ee
The chemical potentials are defined by\footnote{We would like to stress that this is a gauge invariant definition.
The standard boundary condition on the gauge field in AdS/CFT fixes the value of $a_M(u\rightarrow\infty)$.
In this case only the transformations that vanish at $u\rightarrow\infty$ are gauged in the bulk. 
In particular, these transformations do not change the asymptotic value of $a_0$.}
\be
\mu_V = a_0^V(u\rightarrow\infty) \;\;\;\; \mbox{and} \;\;\;\; \mu_A = a_0^A(u\rightarrow\infty) \,.
\ee
In our conventions quarks carry one unit of vector charge and baryons carry $N_c$ units.
Nevertheless we will refer to the vector current as the ``baryon number current".
We are also interested in studying the effects of background ``electromagnetic" fields that couple to
this current, which correspond to turning on spacetime dependent boundary values
of the worldvolume gauge field,
in particular
\be
e_i = f_{0i}(u\rightarrow\infty) \; , \; b_i = \epsilon_{ijk} f_{jk}(u\rightarrow\infty) \,.
\ee

In some situations one may be required to add boundary terms to the action.
These are especially relevant if there is a CS term in the bulk.
In deriving the equations of motion from the variational principle one usually assumes that the surface terms vanish.
However in some instances one has to be more careful.
The surface terms (in the $a_u=0$ gauge) are given in general by
\be
\delta S\Big|_{on-shell} = \int d^4x \, 
\frac{\partial{\cal L}}{\partial a'_\nu} \,
\delta a_\nu \Big|_{u_{min}}^\infty +
\int d^3x \, du \, 
\frac{\partial{\cal L}}{\partial (\partial_\mu a_\nu)} \,
\delta a_\nu\Big|_{x_\mu\rightarrow -\infty}^{x_\mu\rightarrow \infty} \,.
\ee
In holography the boundary values of the fields at $u\rightarrow\infty$ are fixed, so $\delta a_\mu(u\rightarrow\infty) = 0$.
However $\delta a_\mu(u_{min})$ and $\delta a_\mu(x_\mu\rightarrow \pm\infty)$ need not vanish.
Therefore a surface term may be non-trivial if the fields extend to these boundaries.
In order to have a well-defined variational principle we must therefore add boundary 
terms $S_{\partial}(u_{min})+S_{\partial}(x_\mu\rightarrow\pm\infty)$, whose variation cancels the surface terms
in the variation of the bulk action.

These boundary terms also allow one to derive an alternative and useful definition of the conserved currents.
We do this by varying the {\em off-shell} action, now allowing $a_\mu(u\rightarrow\infty)$ to vary,
and then going on-shell by applying the equations of motion.
Due to the boundary terms, only the surface term at $u\rightarrow\infty$ remains. Thus
\be
\label{alternative_current}
j^\mu =  \frac{1}{{\cal N}}\, \frac{\partial{\cal L}}{\partial a'_\mu}(u\rightarrow\infty)\Big|_{on-shell} \,.
\ee
In particular this relates the charge density in the boundary theory to the bulk radial electric field.
The boundary term $S_{\partial}(u_{min})$ for $a_0$ should then be interpreted as a source term for this field.
One could in principle also add boundary terms at $u\rightarrow\infty$.
These have no effect on the derivation of the equations of motion, but may change the value of the on-shell action,
and therefore may lead to additional contributions to (\ref{alternative_current}).

\begin{figure}[htbp]
\begin{center}
\begin{tabular}{ccccc}
\epsfig{file=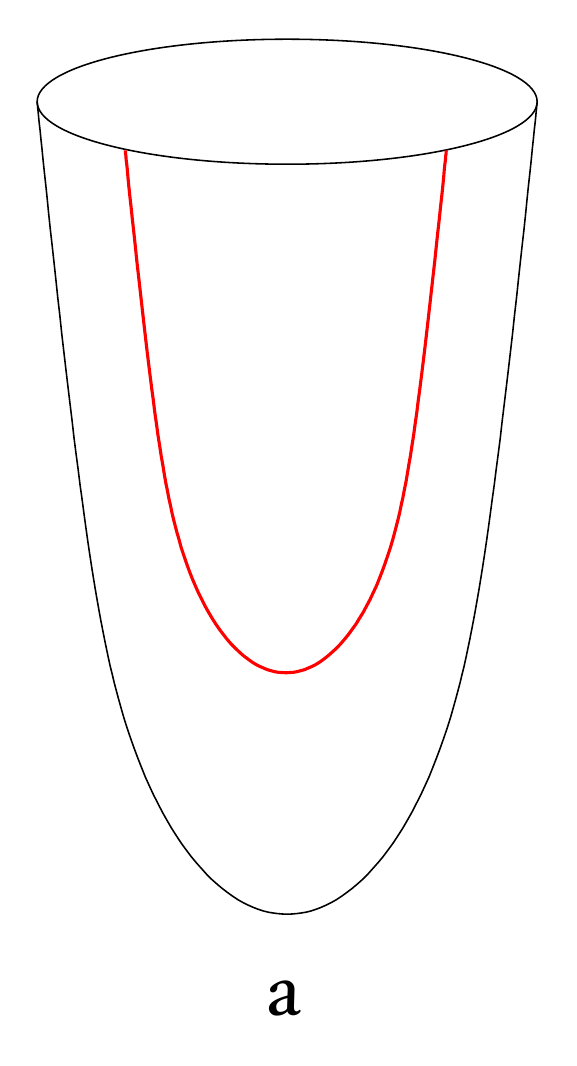,scale=0.3} \;\;\;\;\; &
\epsfig{file=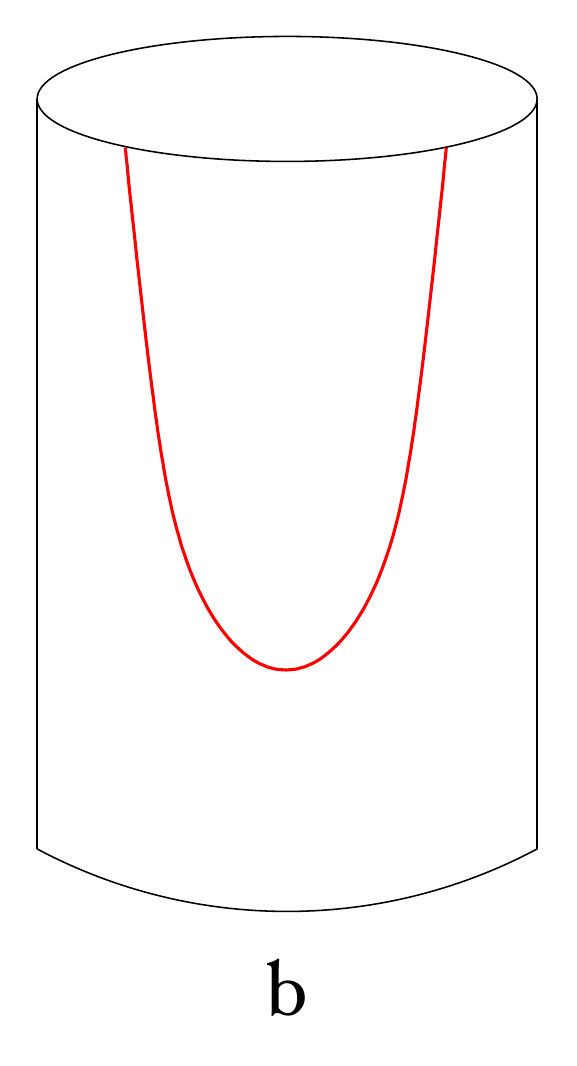,scale=0.3} \;\;\;\;\; &
\epsfig{file=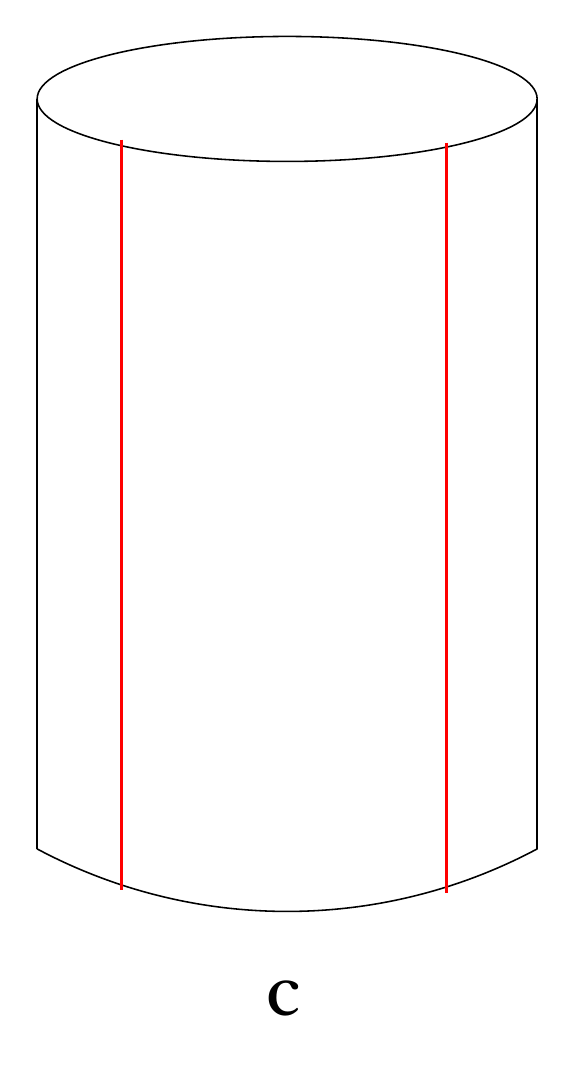,scale=0.3}\;\;\;\;\; &
\epsfig{file=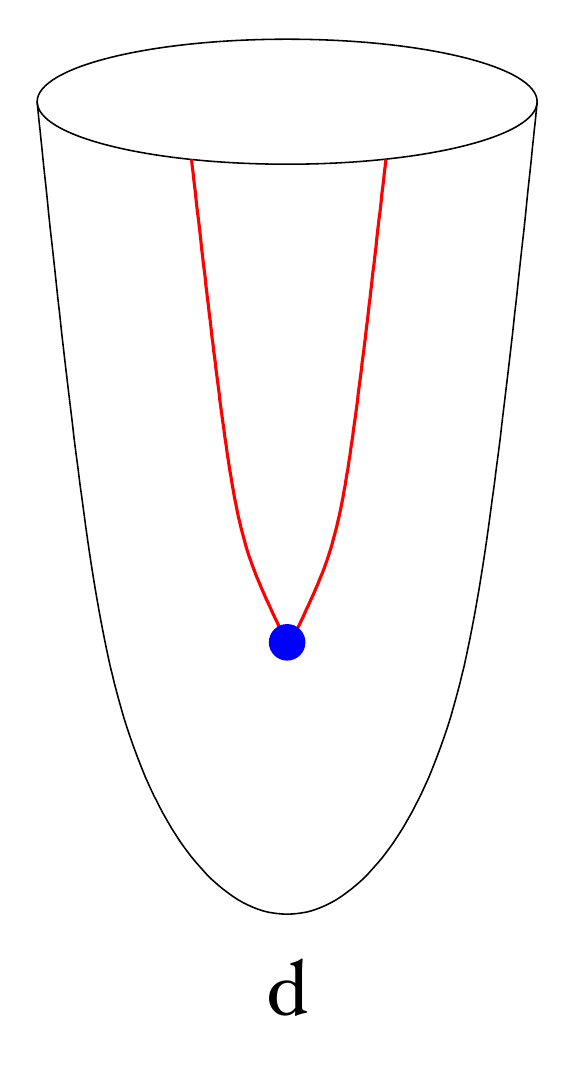,scale=0.3}\;\;\;\;\; &
\epsfig{file=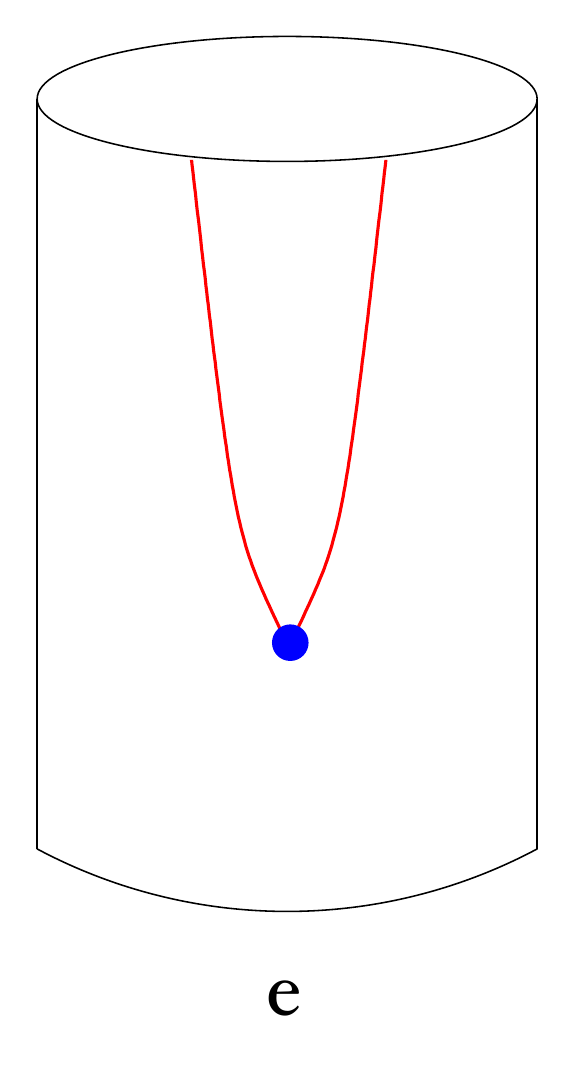, scale=0.3}
\end{tabular}
\caption{D8-brane embeddings in the Sakai-Sugimoto model:
(a) confined vaccum (b) deconfined vacuum
(c) deconfined plasma (d) confined nuclear matter (e) deconfined nuclear matter.}
\label{embeddings}
\end{center}
\end{figure}

A state with a non-zero baryon number density corresponds to an embedding with a radial electric field.
In particular, for a U embedding (in both the confining and deconfining backgrounds) this requires
the addition of a baryonic source at the tip, corresponding to a uniform spatial distribution of wrapped D4-branes
(Fig.\ref{embeddings}d,e) \cite{Bergman:2007wp} (see also \cite{Davis:2007ka,Rozali:2007rx}).
We assume that the distribution is dilute enough so that we can ignore interactions between the D4-branes.
We should therefore include the D4-brane action, which is given by (in the confining background)
\be
S_{D4} =  - {\cal N} V_4 n_{D4} N_c \left(\frac{1}{3} u_0 - a_0^V(u_0)\right) \,,
\ee
where $n_{D4}$ is the density of D4-branes.
The first term is the D4-brane DBI action, and corresponds to the baryon mass,
and the second term comes from the $N_c$ strings that connect each D4-brane to the D8-brane.
The second term is precisely the boundary term at $u_{min}=u_0$ that was discussed above.

The resulting asymptotic behavior of the gauge field is 
\be
\label{asymptotic_a0}
a_0^V(u) \approx 
\mu_V - {2\over 3} {d\over u^{3/2}} \,,
\ee
where $d=N_c n_{D4}$ is the baryon number density.
On the other hand, extremizing the action with respect to $n_{D4}$ fixes the value 
of the gauge field at the tip to $a^V_0(u_0) = u_0/3 = m_{baryon}/N_c$.
This implies, as expected, that a non-zero density configuration exists only when the 
chemical potential is above the baryon mass.
In fact the non-zero density state is always the dominant one.
The transition to ``nuclear matter" occurs at $\mu_V = m_{baryon}/N_c$.
Near the critical point the density scales linearly with the chemical potential $d\sim \mu_V - m_{baryon}/N_c$.
The D4-brane action also sources the embedding field $x_4(u)$, creating a cusp at $u=u_0$.
This can be understood in terms of a force balance condition between the D4-branes
pulling down and the D8-brane pulling up.

At high temperature the preferred embedding is parallel and there is a different finite density solution.
In this case the gauge fields $a_\mu$ and $\bar{a}_\mu$ are independent, and the field theory has
a conserved axial current as well as a baryon current.
For a baryonic solution we take $a^V_0(\infty)=\mu_V$ and $a^A_0(\infty)=0$.
In addition, since the D8-brane and anti-D8-brane reach the horizon we must impose $a_0^V(u_T) = a_0^A(u_T)=0$.
Therefore the radial vector electric field, and thus the baryon number density, is non-zero when $\mu_V>0$.
In this phase $d\sim T^3\mu_V$ for small $\mu_V$.

\subsection{Magnetic catalysis of chiral symmetry breaking}

A strong magnetic field in QCD is believed to catalyze the spontaneous breaking of chiral symmetry
\cite{Gusynin:1994re,Gusynin:1994xp,Miransky:2002rp}. The basic mechanism for this is that in a strong magnetic field
all the quarks sit in the lowest Landau level, and the dynamics is effectively 1+1 dimensional. 
This phenomenon has been exhibited in the
Sakai-Sugimoto model in \cite{Bergman:2008sg,Johnson:2008vna}.\footnote{At non-zero baryon number density  
the magnetic field can actually induce an inverse magnetic catalysis in this model \cite{Preis:2010cq,Preis:2011sp}.}

\begin{figure}[htbp]
\begin{center}
\epsfig{file=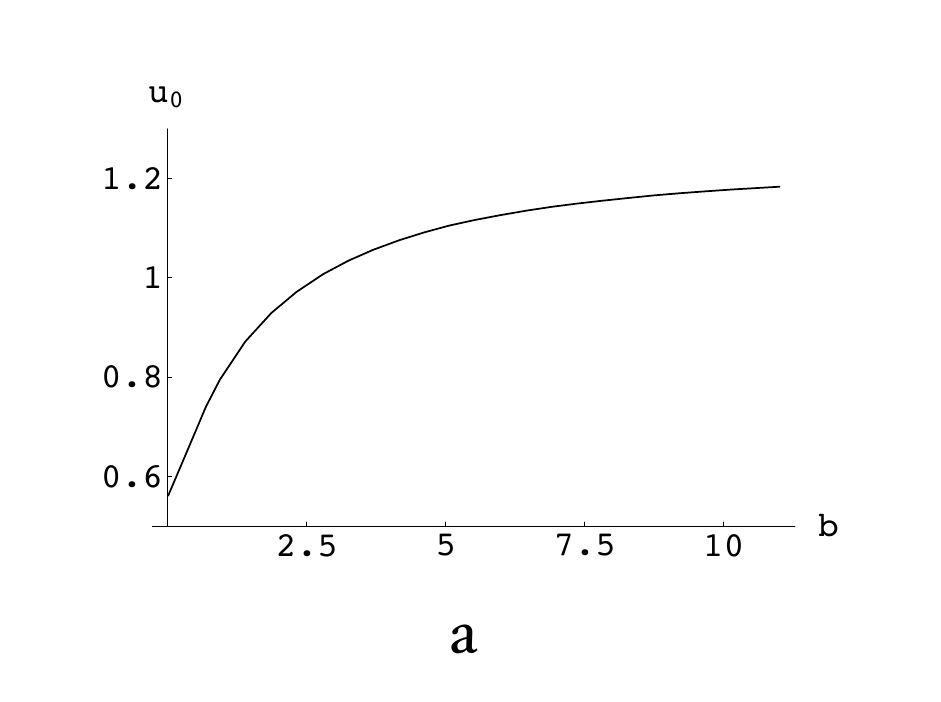,scale=0.55}
\hskip 0.5cm
\epsfig{file=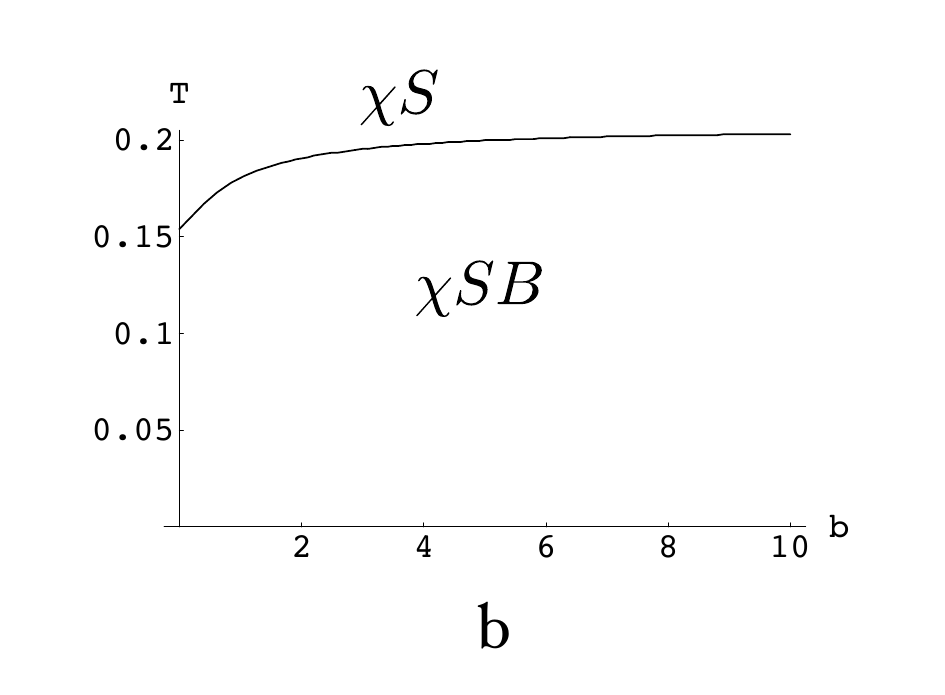, scale=0.55}
\caption{Magnetic catalysis: (a) Chiral symmetry breaking mass scale. (b) Phase diagram.}
\label{fig:magnetic_catalysis}
\end{center}
\end{figure}

With a uniform background magnetic field $b$, the D8-brane action in the deconfining background becomes
\be 
\label{D8_action_deconfined_magnetic}
S_{D8}^{dec} = 
- {\cal{N}}  \int d^4 x \, du \, u^4 \sqrt{\left(f(u) (x_4'(u))^2 + \frac{1}{u^3}\right) \left(1+ \frac{b^2}{u^3} \right)} \,.
\ee
The U embedding has the same form as before (\ref{U_embedding}), 
but now $u_0$
depends on the magnetic field, as shown in Fig.~\ref{fig:magnetic_catalysis}a.
The mass scale associated with chiral symmetry breaking is seen to increase with the magnetic field.
One therefore expects that chiral symmetry breaking becomes more favored as the magnetic field
increases. This is indeed the case, as can be seen by comparing the Euclidean actions of the U and 
parallel embeddings as the temperature and magnetic field are varied. The resulting phase diagram is shown in 
Fig.~\ref{fig:magnetic_catalysis}b.
We observe that in this model the critical temperature approaches a finite value at infinite magnetic field.

A qualitatively similar effect was observed in the D3-D7 model above.
However in the D3-D7 model there is a critical value of $B/T^2$ above which the chiral symmetry
is always broken, whereas in the D4-D8 there is a critical temperature above which the chiral symmetry
is always broken.

It is also instructive to study the effect of the background magnetic field on the mesons.
This was partly done in \cite{Johnson:2009ev}, in which the high spin mesons were studied. 
It was shown that the magnetic field enhances their stability by increasing their angular momentum, and thereby
increasing the dissociation temperature at which they fall apart into their quark constituents. 
This is consistent with the above results.

\subsection{Anomalous currents}
\label{anomalous_currents_section}

The chiral anomaly leads to two interesting phenomena when both the magnetic field and chemical
potential are non-zero.
The first is the generation of anomalous currents in the chiral symmetric phase of QCD.
In \cite{Metlitski:2005pr,Newman:2005as} it was shown that 
the combination of a magnetic field and a non-zero baryon chemical potential generates an axial current 
\be 
\vec{J}_A = \frac{e}{2\pi^2} \mu^{phys}_B \vec{B} \,.
\ee
Since the source for this current is the anomaly it is an exact result, and should be valid
in particular at strong coupling.
Similarly, an anomalous vector current is generated in a non-zero axial chemical potential: 
\be
\vec{J}_V = \frac{e}{2\pi^2} \mu^{phys}_A \vec{B} \,.
\ee
This is known as the ``chiral magnetic effect", and may have some relevance to heavy ion physics at RHIC
\cite{Fukushima:2008xe}.
Within the D3/D7 model, this effect has been discussed in \cite{Hoyos:2011us}.

In the Sakai-Sugimoto model, the chiral-symmetric phase corresponds to the parallel 
D8-$\overline{\mbox{D8}}$ embedding
in the deconfined background,
and the chiral anomaly is encoded in the five-dimensional CS term (\ref{CS_action}).
The background magnetic field and chemical potentials correspond to
different components of the worldvolume gauge field. Through the five-dimensional CS term
these source a third component, which corresponds to a current in the four-dimensional theory
\cite{Bergman:2008qv}.
Let us review the calculation of the anomalous axial current in the Sakai-Sugimoto model.
The calculation of the anomalous vector current is virtually identical.

To be specific, we will consider a background magnetic field in the $x_1$ direction
by turning on a background gauge field $a_3^V =x_2 b$.
A non-trivial boundary value of $a_0^V$ will then source, via the CS term, a non-trivial $a_1^A$.
In general $a_0^V$ and $a_1^A$ can depend on both $u$ and $x_2$ in this case, although on-shell they will
depend only on $u$.
We will take $a_0^V(\infty)=\mu_V$ and $a_1^A(\infty)=0$.
The D8-brane DBI and CS actions in this case become
\be 
S_{DBI}^{dec} &=&  - {\cal{N}} \int_{u_T}^\infty d^4 x \, du \, u^{5/2} \sqrt{\Big(1
-(a_0^{V\prime})^2 + f(u)(a_1^{A\prime})^2 \Big)\left(1+{b^2\over u^3}\right)} \\[5pt]
S_{CS} & = &    - {\cal N} 
\int d^4 x\, du \, \left[b (a_0^V a_1^{A\prime} - a_0^{V\prime} a_1^A)
+ a_{3}^{V}(  a_0^{V\prime} \partial_2 a_1^A - \partial_2 a_0^V a_1^{A\prime} )\right] .
\ee
The necessary boundary terms are
\be
\label{boundary_action}
S_{\partial} = - \frac{1}{2}{\cal N}  \left.\int d^3 x \, du\, a_3^V (a_0^V a_1^{A\prime} - a_0^{V\prime} a_1^A)
\right|_{x_2\rightarrow -\infty}^{x_2\rightarrow\infty} \,.
\ee
By integrating by parts the last two terms in the CS action one can show that 
up to a surface term at $u\rightarrow\infty$,
the bulk CS and boundary actions combine into a bulk action
\be
\label{CS+boundary}
S_{CS} + S_{\partial} =  - {\cal N} \int d^4 x \, du \, \left[\frac{3}{2} b (a_0^V a_1^{A\prime} - a_0^{V\prime} a_1^A) 
- \frac{1}{2} a_3^{V\prime}(a_0^V\partial_2 a_1^A - \partial_2 a_0^V a_1^A)\right] .
\ee
One can get rid of the remaining surface term by adding a boundary term at $u\rightarrow\infty$
\cite{Bergman:2008qv}, however this particular term does not contribute to the on-shell action,
so we might as well ignore it.\footnote{Other boundary terms at $u\rightarrow \infty$ could affect
the on-shell action, and therefore the currents. See for example \cite{Rebhan:2009vc,Rebhan:2010ax}.}

The equations of motion for $a_0^V(u)$ and $a_1^A(u)$ can be integrated once to yield
\be
\label{EOMa0dec}
\frac{\sqrt{u^5+b^2 u^2}\, a_0^{V\prime}(u)}{\sqrt{1 - (a_0^{V\prime}(u))^2 + f(u)(a_1^{A\prime}(u))^2}}
&=& - 3ba_1^A(u) + d\\[5pt]
\label{EOMa1dec}
\frac{\sqrt{u^5+b^2 u^2}\, f(u) a_1^{A\prime}(u)}{\sqrt{1 - (a_0^{V\prime}(u))^2 + f(u)(a_1^{A\prime}(u))^2}}
&=& - 3ba_0^V(u)\,,
\ee
where $d$ is the baryon number charge density. The integration constant in the $a_1^A$ equation
vanishes
since $a_0^V(u_T)=0$ and $f(u_T)=0$.
Using (\ref{alternative_current}) and (\ref{EOMa1dec}) we can then evaluate the axial current:
\be 
j_A^1 = \frac{3}{2} b \, a_0^V(\infty)=  \frac{3}{2} b \mu_V \,.
\ee

The correctly normalized physical currents are given by $J = 2(2\pi\alpha'{\cal N}/R^5)j$,
where the factor of 2 comes from adding the anti-D8-brane contribution,
and the physical chemical potentials are $\mu^{phys} = (R/(2\pi\alpha'))\mu$.
Thus in terms of the physical variables our result translates to
\be 
\vec{J}_A =  \frac{N_c}{4\pi^2} \mu^{phys}_V \vec{B} \,.
\ee
Similarly, for an axial chemical potential we would get
\be 
\vec{J}_V =  \frac{N_c}{4\pi^2} \mu^{phys}_A \vec{B} \,.
\ee
Interestingly, our results are half of the weak-coupling results (where $e=N_c$ in the holographic model).
It has been argued that the discrepency in the axial current is due to a different treatment of the triangle
anomaly, consistent vs. covariant, and can be corrected by adding an appropriate Bardeen counterterm on the 
boundary \cite{Rebhan:2010ax}. However, the same counterterm leads to a vanishing vector current.
This issue is still under investigation.

\subsection{The pion gradient phase}
\label{pion_gradient_section}

In the broken chiral symmetry phase the chiral anomaly leads to a novel finite density phase 
that dominates over nuclear matter at large magnetic fields \cite{Son:2007ny}.
In this phase the baryon charge is carried not by baryons but rather by a non-zero pion gradient background:
\be
\label{pion_gradient_baryon_charge}
D = {e\over 4\pi^2 f_\pi} \vec{B} \cdot \vec{\nabla}\pi^0 \,,
\ee
where
\be
\label{pion_gradient}
\vec{\nabla}\pi^0 = {e\over 4\pi^2 f_\pi} \mu^{phys}_B \vec{B} \,.
\ee

In the Sakai-Sugimoto model (in the $a_u=0$ gauge) the 
pseudoscalar meson appears in the zero mode of $a_\mu^A$ (\ref{pseudoscalar}), so
\be
\partial_\mu\varphi(x^\mu) = - a^A_\mu(x^\mu,u\rightarrow\infty) \,.
\ee
As in the chiral-symmetric phase, the presence of a vector chemical potential together
with a background magnetic field sources a component of the axial gauge field,
which in this case corresponds to a non-trivial gradient of the pseudoscalar field \cite{Bergman:2008qv,Thompson:2008qw}.
Since there is only one flavor this field should really be thought as the $\eta'$ meson.
For simplicity, we will consider only the confined phase with the anti-podal D8-brane embedding,
namely $u_0=u_{KK}$.
(The results do not change qualitatively for more general U-shape embeddings,
or for U-shape embeddings in the deconfined phase.)

Following \cite{Bergman:2008qv}, the D8-brane DBI action in this case is
\be
\label{DBI_pion_gradient}
S_{DBI}^{con} =   - {\cal{N}} \int_{u_{KK}}^\infty d^4 x \, du \, u^{5/2} \sqrt{\left(\frac{1}{f(u)}
-(a_0^{V\prime})^2 +(a_1^{A\prime})^2 \right)\left(1+{b^2\over u^3}\right)}\,,
\ee
and the CS plus boundary actions are the same as in the deconfined phase (\ref{CS+boundary}).
The equations of motion integrate to
\be
\label{EOMa0con}
{\sqrt{u^5+ b^2 u^2}\, a_0^{V\prime}(u) 
\over\sqrt{{1\over f(u)} - (a_0^{V\prime}(u))^2 + (a_1^{A\prime}(u))^2}}
&=& - 3ba_1^A(u) + N_c n_{D4} \\[5pt]
\label{EOMa1con}
{\sqrt{u^5+ b^2 u^2}\, a_1^{A\prime}(u) 
\over\sqrt{{1\over f(u)} - (a_0^{V\prime}(u))^2 + (a_1^{A\prime}(u))^2}}
&=& - 3ba_0^V(u) + c \,,
\ee
where we have explicitly included the baryon sources in the $a_0^V$ equation.
Our boundary conditions are now $a_0^V(\infty)=\mu_V$ and $a_1^A(\infty) = -\nabla\varphi(x^\mu)$.
In particular $a_1^A(\infty)$ is a field rather than a parameter in the boundary theory, and
we must minimize the action with respect to its value. This simply sets $j_A^1=0$ and 
therefore sets the integration constant in the $a_1^A$ equation to  $c=\frac{3}{2} b\mu_V$.\footnote{This is also consistent 
with the fact that there are no quarks in this phase to carry such a current.}

We can now compute the total baryon number charge density $d$ using the same procedure as in the previous section for the current.
In the absence of sources $n_{D4}=0$ and we find
\be
\label{density_pion_gradient}
d = - \frac{3}{2}b\, a_1^A(\infty) = \frac{3}{2}b\, \nabla\varphi \,.
\ee
Let us express this in terms of the physical variables.
First we must define a field with a  canonically normalized kinetic term.
Inserting (\ref{pseudoscalar}) into the action (\ref{DBI_pion_gradient}) we find that the
canonically normalized field is given by
\be
\eta'(x^\mu) = \frac{R^2}{2\pi\alpha'}f_{\eta'} \varphi(x^\mu) \; , \;\; f_{\eta'}^2 
= \frac{N_c u_{KK}^{3/2}}{4\pi^4\alpha'} \,.
\ee
Converting to physical variables we then find
\be
D = \frac{N_c}{4\pi^2 f_{\eta'}}\, \vec{B}\cdot\vec{\nabla}\eta' \,,
\ee
in agreement with (\ref{pion_gradient_baryon_charge}).
We would like to stress that this agreement did not depend on the specific value of $f_{\eta'}$
required for canonical normalization, since it cancels out when we express the result in terms of $\varphi$.
The correct numerical factor of $1/(4\pi^2)$ is a direct consequence of including the proper boundary
terms in the action, leading to the ``3/2" in (\ref{density_pion_gradient}).

\begin{figure}[htbp]
\begin{center}
\epsfig{file=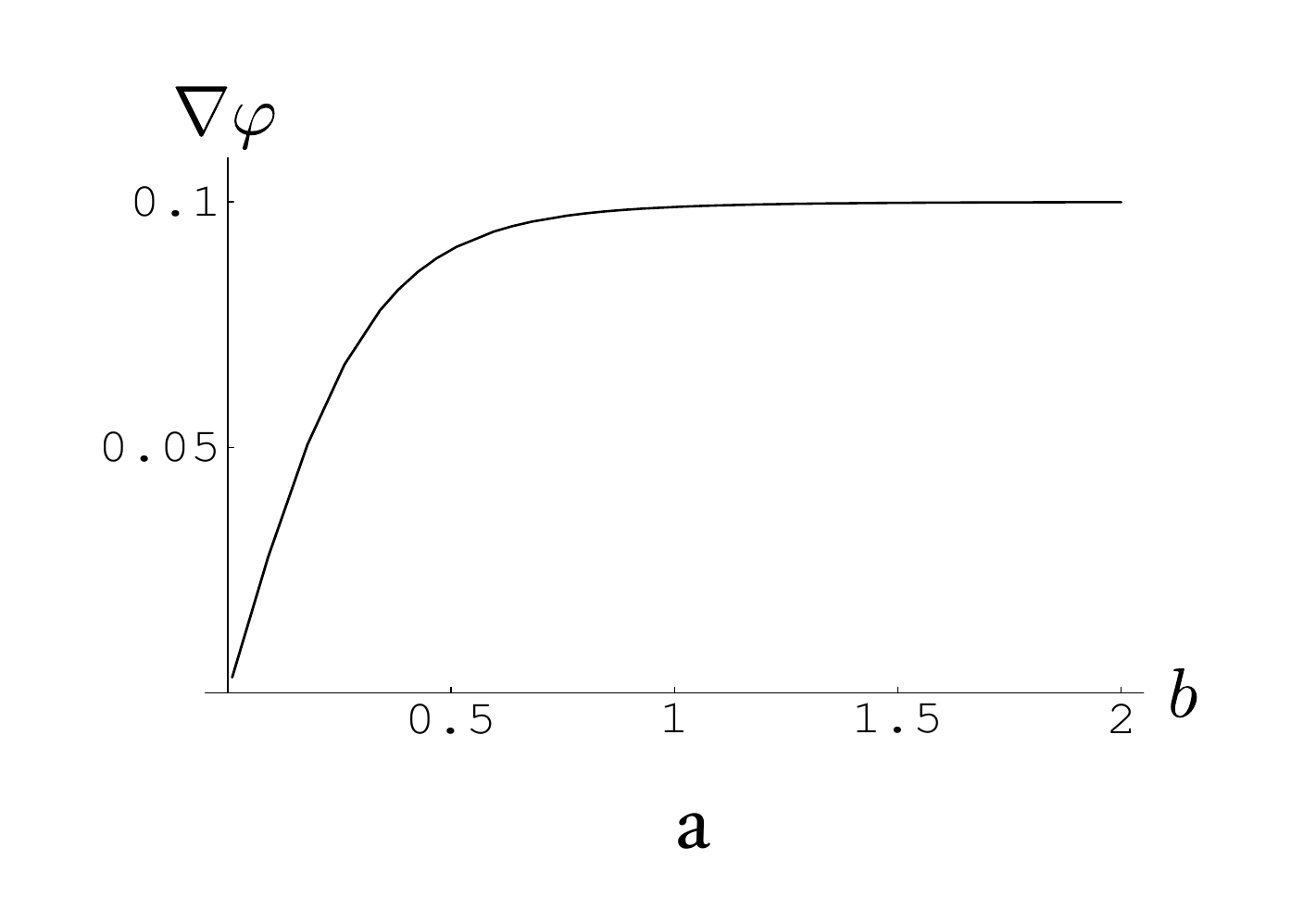,scale=0.37}
\hskip 0.5cm
\epsfig{file=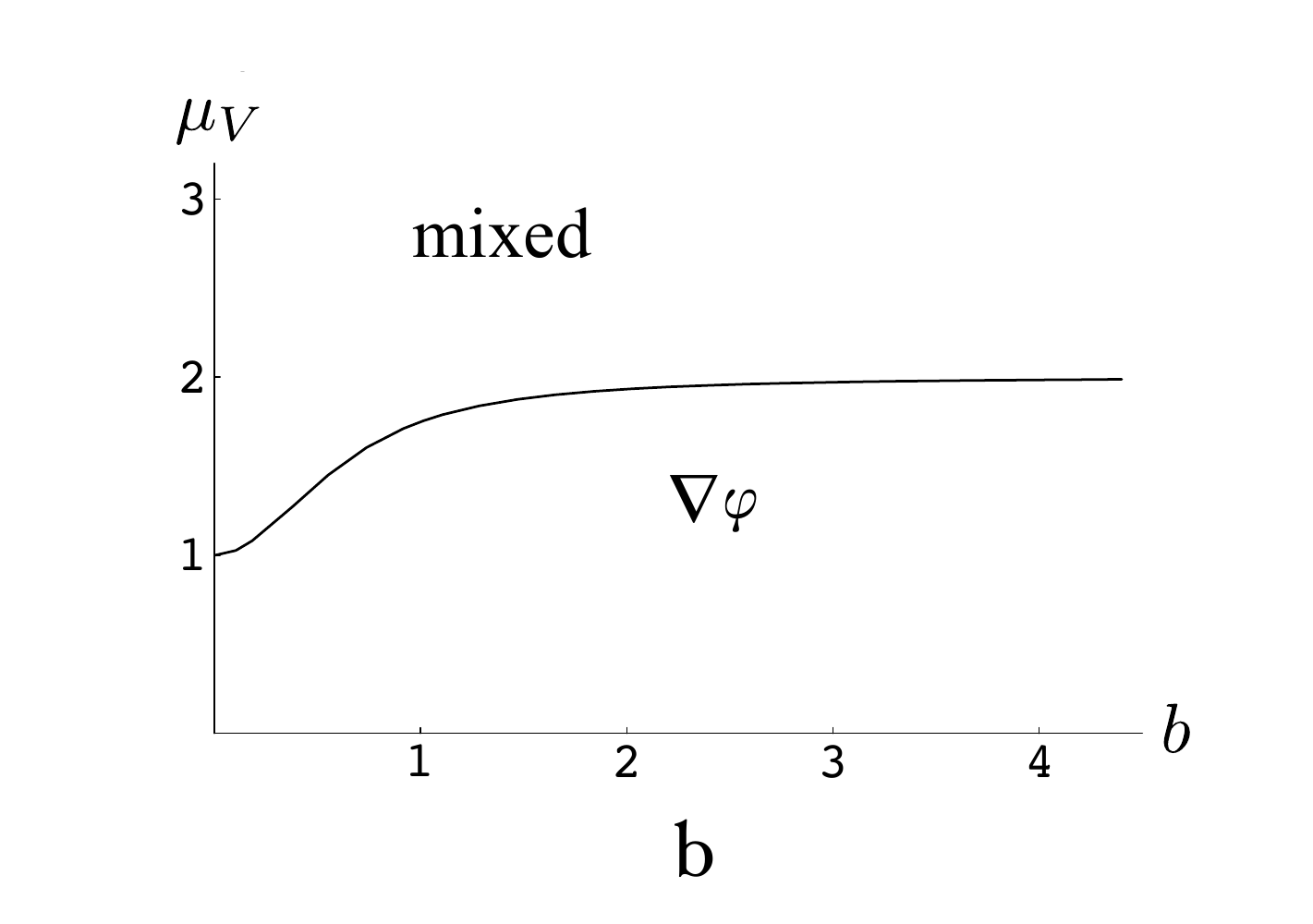,scale=0.37}
\caption{{\bf a.} The pion gradient. {\bf b.} Phase diagram with magnetic field and baryon chemical potential.}
\label{phi_vs_b}
\end{center}
\end{figure}

To find the value of the gradient $\nabla\varphi$ we need to solve (\ref{EOMa0con}) and (\ref{EOMa1con}).
The result will not be as simple as (\ref{pion_gradient}).
In particular it is not linear in the magnetic field, since we are using the full non-linear DBI action.
It turns out that a closed form solution can be found in terms of a new variable
\be
y = \int_{u_{KK}}^u 
{3b \, d\tilde u\over\sqrt{f(\tilde u)}\sqrt{\tilde u^5\left(1+ b^2\, \tilde u^{-3}\right) 
- \left(\frac{3}{2}  b\mu_V\right)^2
+ (3b \nabla\varphi)^2}} \,.
\ee
The solution is
\be
a_0^V(y) = {\mu_V\over 2}\left({\cosh y_{\phantom\infty}\over\cosh y_\infty} + 1\right) \; , \;\;
a_1^A(y) = - {\mu_V\over 2}\, {\sinh y_{\phantom\infty}\over\cosh y_\infty} \,,
\ee
where $y_\infty = y(u\rightarrow\infty)$. The pseudoscalar gradient is then given by
\be 
\nabla\varphi = - a_1^A(\infty) = \frac{\mu_V}{2} \tanh y_\infty \,.
\ee
The dependence on $b$ is shown in Fig.~\ref{phi_vs_b}a.
For small $b$ the behavior is linear in $b$:
\be
\nabla\varphi \approx \frac{\pi}{2u_{KK}^{3/2}}\mu_V b \,,
\ee
and in terms of the physical quantities:
\be
\vec{\nabla}\eta' \approx \frac{N_c}{4\pi^2 f_{\eta'}} \mu_V^{phys} \vec{B} \,,
\ee
in agreement with the single flavor version of (\ref{pion_gradient}).

As in the case with no magnetic field, an embedding that includes sources is possible
above a critical value of the chemical potential, which then becomes the dominant configuration.
This describes a ``mixed phase" that includes both ``pion-gradient" matter and nuclear matter,
with a total baryon number density
\be
d = \frac{3}{2}b\, \nabla\varphi + N_c n_{D4}\,.
\ee
As before, one can find a closed form solution for $a_0^V$ and $a_1^A$,
and from it determine the values of $\nabla\varphi$ and $n_{D4}$ in terms of the magnetic
field $b$ and baryon number chemical potential $\mu_V$.
The resulting phase diagram is shown in Fig.~\ref{phi_vs_b}b.
In particular, the critical value of $\mu_V$ is determined by setting $n_{D4}=0$.
The relative proportion of baryons in the mixed phase increases with $\mu_V$ and decreases with $b$.

\subsection{Magnetic phase transition}

The high-temperature chiral-symmetric phase of this model also exhibits an interesting magnetic phenomenon
associated with the distribution of the baryonic charge \cite{Lifschytz:2009sz}. In what follows we analyze the situation for the D8-brane.
As in the broken chiral symmetry phase above,
the distribution of baryonic charge along the radial direction $u$ changes with $b$.
We can therefore identify two types of baryonic charge,
one originating from the horizon, and the other from outside the horizon.
The latter $d_*$, corresponds to D4-branes that are radially smeared inside the D8-brane.
This can be best seen from the longitudinal and transverse conductivities \cite{Lifschytz:2009si}
\be
\label{D4D8_long_conductivity}
\sigma_{L} &=& \frac{\sqrt{u_{T}^{8}+b^2u_{T}^{5}+u_{T}^{3}(d-d_{*})^{2}}}{u_{T}^{3}+b^2} \\
\label{D4D8_trans_conductivity}
\sigma_{T} &=& \frac{b(d-d_{*})}{{u_{T}^{3}+b^2}}+\frac{d_{*}}{b} \,,
\ee
where $d_*=-3ba_1(u_T)$.
In particular only the horizon charge $d-d_*$
contributes to the longitudinal conductivity.
In the transverse conductivity, the horizon charge contributes as an ordinary dissipative fluid,
whereas the charge outside the horizon $d_*$ behaves as a dissipation-free fluid.
This is consistent with an interpretation of $d_*$ as the charge filling the lowest Landau level.
As the magnetic field increases more of the charge is ``lifted" from the horizon,
representing the transition to the lowest Landau level in the boundary theory.

\begin{figure}[htbp]
\begin{center}
\begin{tabular}{cc}
\epsfig{file= 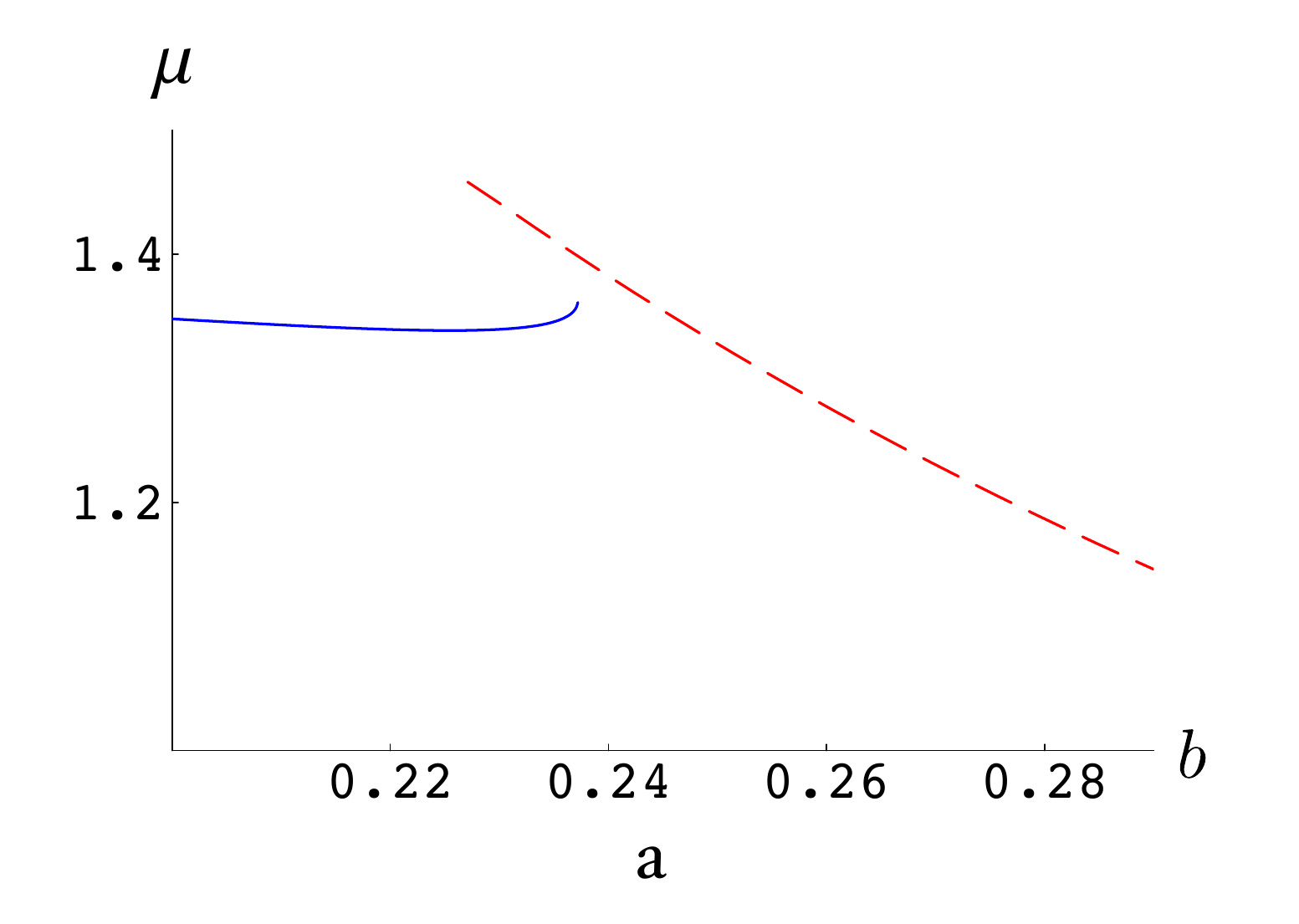,width=5.8cm} &
\epsfig{file= 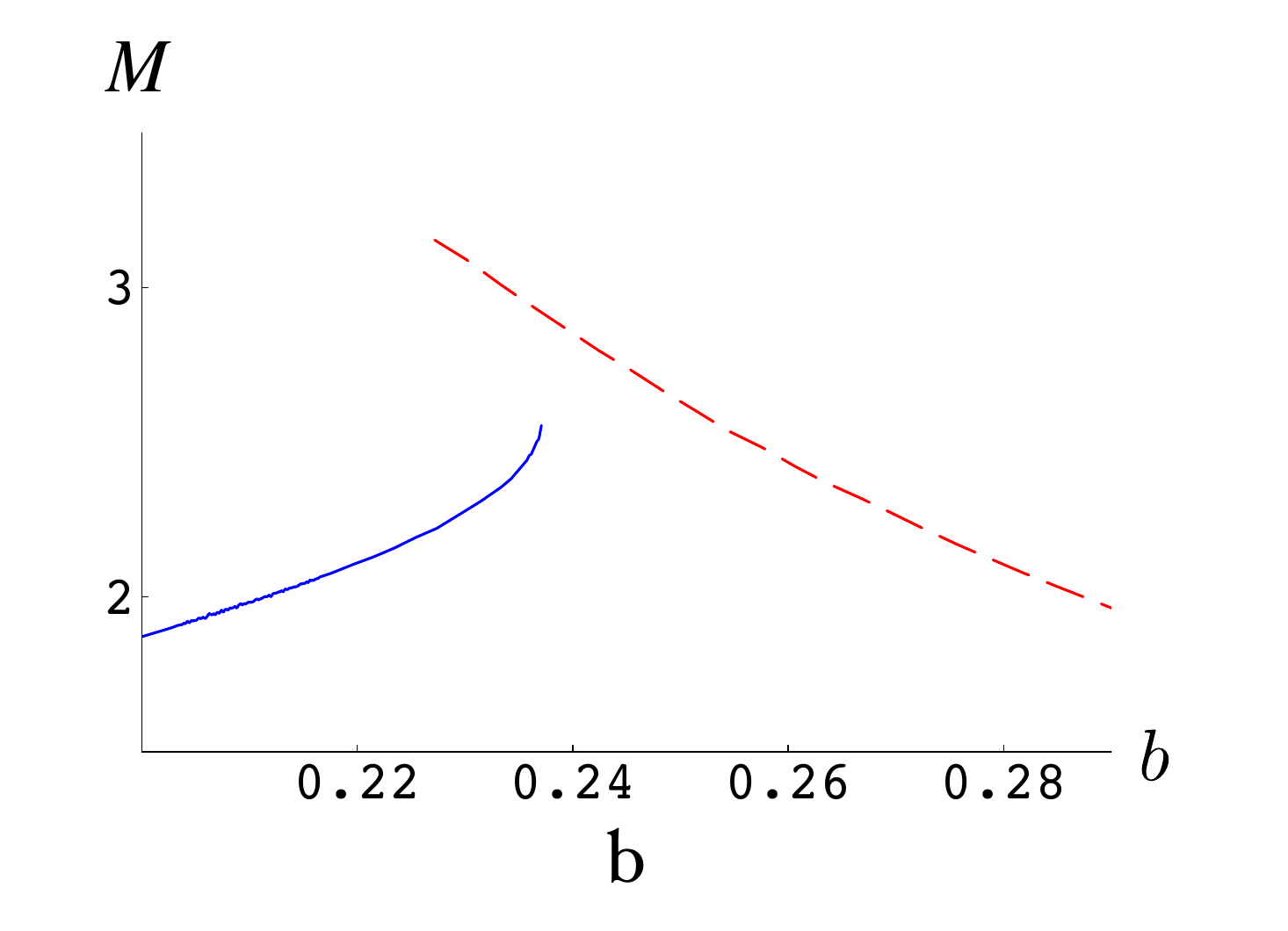,width=5.6cm}
\end{tabular}
\caption{(a) $\mu$ and (b) $M$ as functions of $b$ for $d=1$ and $T=0.09$, below the critical point.  There are now two branches of stable solutions, and the phase transition between them occurs at $b = 0.235$.}
\label{mu_and_M_vs_h_jump_fig}
\end{center}
\end{figure}

In fact for a fixed density at low enough temperature this transition is  a first order phase transition as a function of the magnetic field, in which the charges jump into the lowest Landau level.
This is easiest to see in the zero temperature limit.\footnote{Strictly speaking, at zero temperature the theory 
is in the confining (and broken chiral symmetry) phase. We are considering the meta-stable state 
obtained by adiabatically reducing the temperature.} 
In this case one can solve the gauge field equations analytically
in terms of a variable
\be
\label{zintegralequation}
z=\int_{0}^{u} \frac{3b\, d\tilde{u}}{\sqrt{\tilde{u}^5 +b^2 \tilde{u}^2 + d^{2}\, \cosh^{-2}z_{\infty}}} \,,
\ee
where $z_{\infty}= z(u\rightarrow\infty)$.
The solution is 
\be
a_{0}^V = \frac{d \sinh z}{3b\cosh z_{\infty}} \; , \;\;
a_{1}^A = \frac{d \cosh z}{3b\cosh z_{\infty}}-\frac{d}{3b} \ .
\ee
There are actually three solutions, representing two stable phases and an unstable phase. 
As the magnetic field $b$ is increased, for a fixed total baryon number density $d$, one finds
a first order phase transition between the two stable phases.
Both the magnetization and the chemical potential are discontinuous in this transition (Fig.~\ref{mu_and_M_vs_h_jump_fig}),
which is reminiscent of a metamagnetic phase transition. The large magnetic field phase represents the situation where all the charge is in the lowest Landau level, 
with the chemical potential and free energy given by
\be
\label{muandF}
\mu=\frac{d}{3b} \ \ ,\ \ F=\frac{d^2}{6b} \ .
\ee

\begin{figure}[htbp]
\begin{center}
\begin{tabular}{cc}
\epsfig{file=  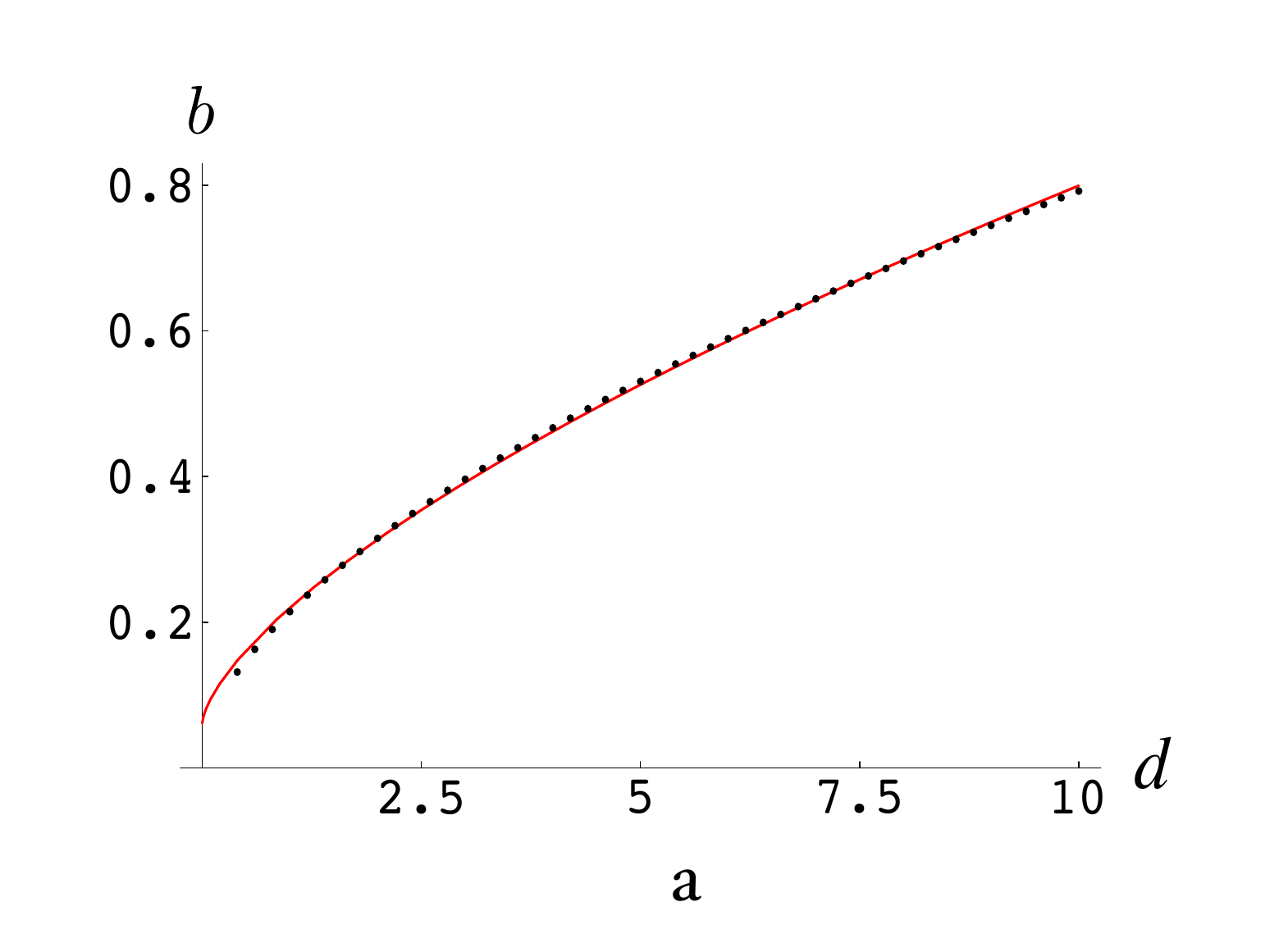,width=5.4cm} &
\epsfig{file= 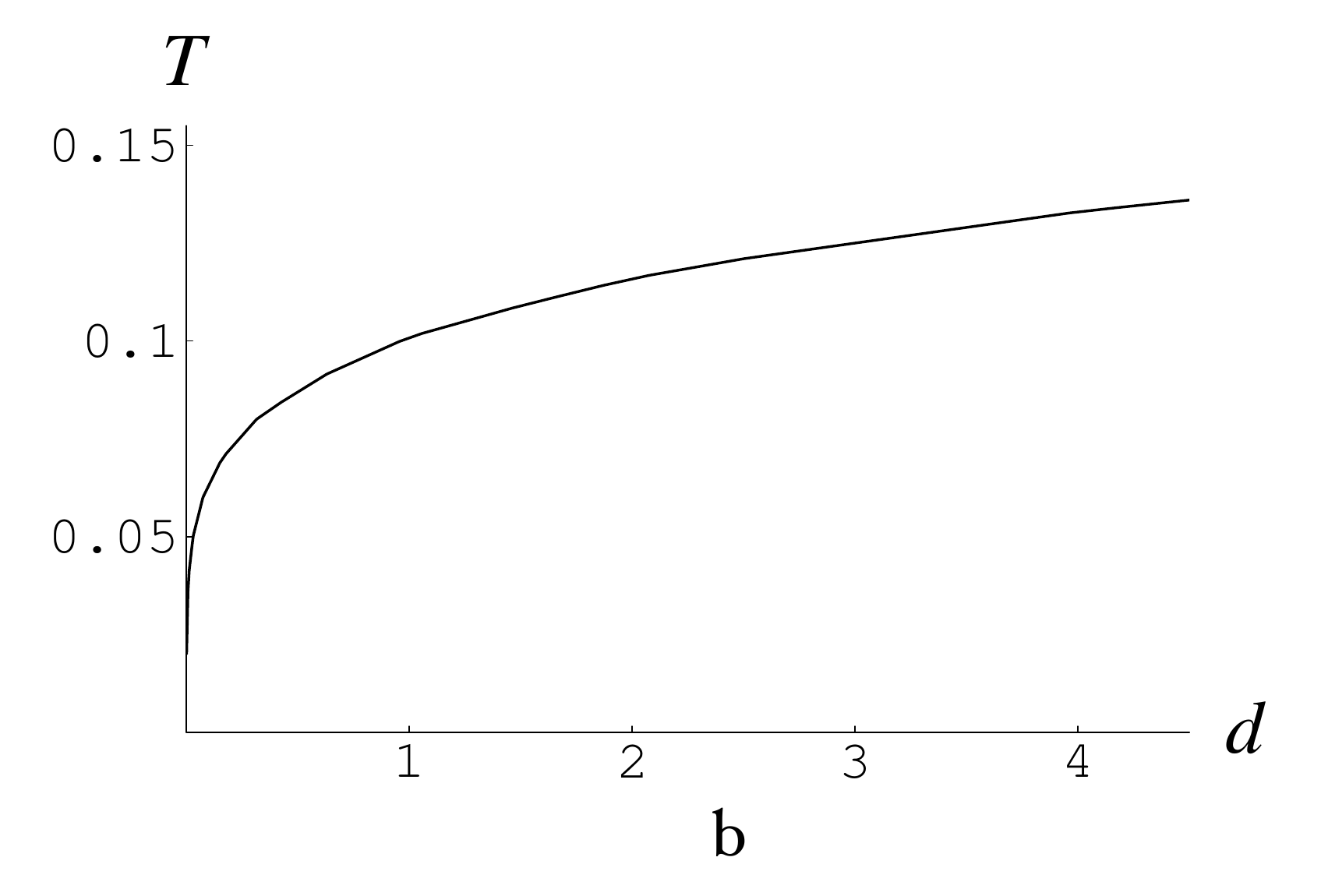,width=5.4cm}
\end{tabular}
\caption{(a) The phase diagram in the $d$-$b$ plane at $T=0.07$, and (b) the critical line in the $T$-$d$ plane.}
\label{d-T_critical_line_fig}
\end{center}
\end{figure}

The magnetic transition persists also at non-zero temperatures that are low relative to the density $d$.
Too a very good approximation this happens when $b \sim d^{2/3}$
(Fig.~\ref{d-T_critical_line_fig}a), which is the behavior expected for the lowest  Landau level.
At high temperature the transition disappears (Fig.~\ref{d-T_critical_line_fig}b).

\section{The D3-D7' model}

The study of magnetic properties of planar matter is a very active area of research in condensed matter physics.
A simple holographic model for charged fermions in three dimensions can be obtained by T-dualizing the 
D4-D8 setup.
This leads to a D3-D7 configuration with the two sets of branes intersecting on a plane
\cite{Rey_strings,Davis:2008nv}
(for a related model using a D2-D8 system see \cite{Jokela:2011eb}).
However, unlike in the D4-D8 configuration, here the branes have a mutually transverse coordinate.
On the one hand this allows the fermions to be massive, but on the other hand it leads to an instability since the different
branes repel.

\subsection{Stable embeddings}

First we have to address the issue of stability.
As before, we will employ the probe approximation and consider a single probe D7-brane.
The background (at finite temperature) in this case is 
\begin{eqnarray}
\label{D3metric}
 L^{-2} ds_{10}^2 &=& r^2 \left(-h(r)dt^{2}+dx^2+dy^2+dz^2\right)+
 r^{-2} \left(\frac{dr^2}{h(r)}+r^2 d\Omega_5^2\right) \\
\label{RR_5-form}
F_5 &=& 4L^4\left(r^3 dt\wedge dx\wedge dy\wedge dz\wedge dr 
+  d\Omega_5 \right) \,,
\end{eqnarray}
where $h(r)=1-r_T^4/r^4$, $r_T=\pi L T$ and $L^2=\sqrt{4\pi g_{s} N_c}\, \alpha'$. 
It is convenient to parameterize the five-sphere as an $S^2\times S^2$ fibered over an interval:
\be
 d\Omega_5^2 = d\psi^2 + 
 \cos^2\psi (d\Omega_2^{(1)})^2 + \sin^2\psi (d\Omega_2^{(2)})^2 \,,
\ee
where $0\leq \psi \leq \pi/2$. 
The first $S^2$ shrinks at the ``south pole" $\psi=\pi/2$ and the second $S^2$ at the ``north pole" $\psi=0$.
The D7-brane wraps the two $S^2$'s and extends along $(x,y)$, and has an embedding described
by $z(u)$ and $\psi(u)$.
In particular $\psi$ is dual to the fermion bi-linear operator in the field theory corresponding to the fermion mass.
The D7-brane DBI action in this background is given by
\be
 S_{DBI}  =  
 -4{\cal N} \int d^3 x \, dr\, r^2 \cos^2\psi \sin^2\psi
  \sqrt{1+r^4h(r) z'^2+r^2 h(r)\psi'^2} \,,
\ee
where ${\cal N} \equiv 4\pi^2 \mu_7 L^8 /g_s$. 
A massless embedding would correspond to $\psi = \pi/4$. However the fluctuations contain a mode that 
violates the Breitenlohner-Freedman bound, and therefore the embedding is unstable.
This can also be seen by trying a more general embedding with a large $r$ behavior of
\be
\label{ansatz}
\psi(r) \sim {\pi\over 4} + cr^\Delta \,.
\ee
The equation of motion for $\psi$ gives $\Delta(\Delta + 3) = -8$, which does not have a real solution.

Fortunately, the D7-brane can be stabilized by turning on some worldvolume flux \cite{Myers:2008me},
in this case on the two-spheres \cite{Bergman:2010gm}:
\be
2\pi\alpha' F = {L^2\over 2} \left(f_1 d\Omega_2^{(1)} + f_2 d\Omega_2^{(2)}\right) \; , \;\;
f_i = {2\pi\alpha'\over L^2}\, n_i \;\; (n_i\in \mathbb{Z}) \,.
\ee
This changes the DBI action,
\be
 \lefteqn{S_{DBI} =} \nonumber \\  
 && -{\cal N}\int d^3x \, dr\, r^2\sqrt{\left(4\cos^4\psi + {f_1^2} \right)
 \left(4\sin^4\psi + {f_2^2} \right) \left(1+r^4 h z'^2+r^2 h \psi'^2\right)} ,
\ee
and there is now also a CS term which gives,
\be
S_{CS} = -{\cal N} f_1 f_2 \int d^3x\, dr\, r^4 z'(r) \,.
\ee
The asymptotic behavior of $\psi(r)$ is now
\be
\psi(r) \sim \psi_{\infty} + mr^{\Delta_{+}}-c_{\psi}r^{\Delta_{-}}\,,
\ee
where $\psi_\infty$ is determined by the solution of
\begin{equation}
\label{const_solution}
(f_1^2 + 4\cos^4\psi_\infty)\sin^2\psi_\infty = (f_2^2 + 4\sin^4\psi_\infty)\cos^2\psi_\infty \,,
\end{equation}
and
\be
\Delta_{\pm} = -\frac{3}{2}\pm \frac{1}{2}\sqrt{9+16\frac{f_1^2+16\cos^6\psi_\infty-12\cos^4\psi_\infty}{f_1^2+4\cos^6\psi_\infty}} \ .
\label{deltapm}
\ee
In particular, the embedding is stable for a large enough flux.
The coefficient of the leading term is related to the fermion mass, 
and that of the subleading term corresponds to the bi-linear condensate.
Note that the scaling dimension of the bi-linear operator is given by $-\Delta_-$.
This represents a large anomalous dimension, which is not surprising given that the model is non-supersymmetric.
We should require however that the operator be relevant, namely that $\Delta_-\geq -3$, and therefore that $\Delta_+ \leq 0$,
in order to consider the leading term as a ``mass deformation".

\begin{figure}[htbp]
\begin{center}
\begin{tabular}{cc}
\epsfig{file= 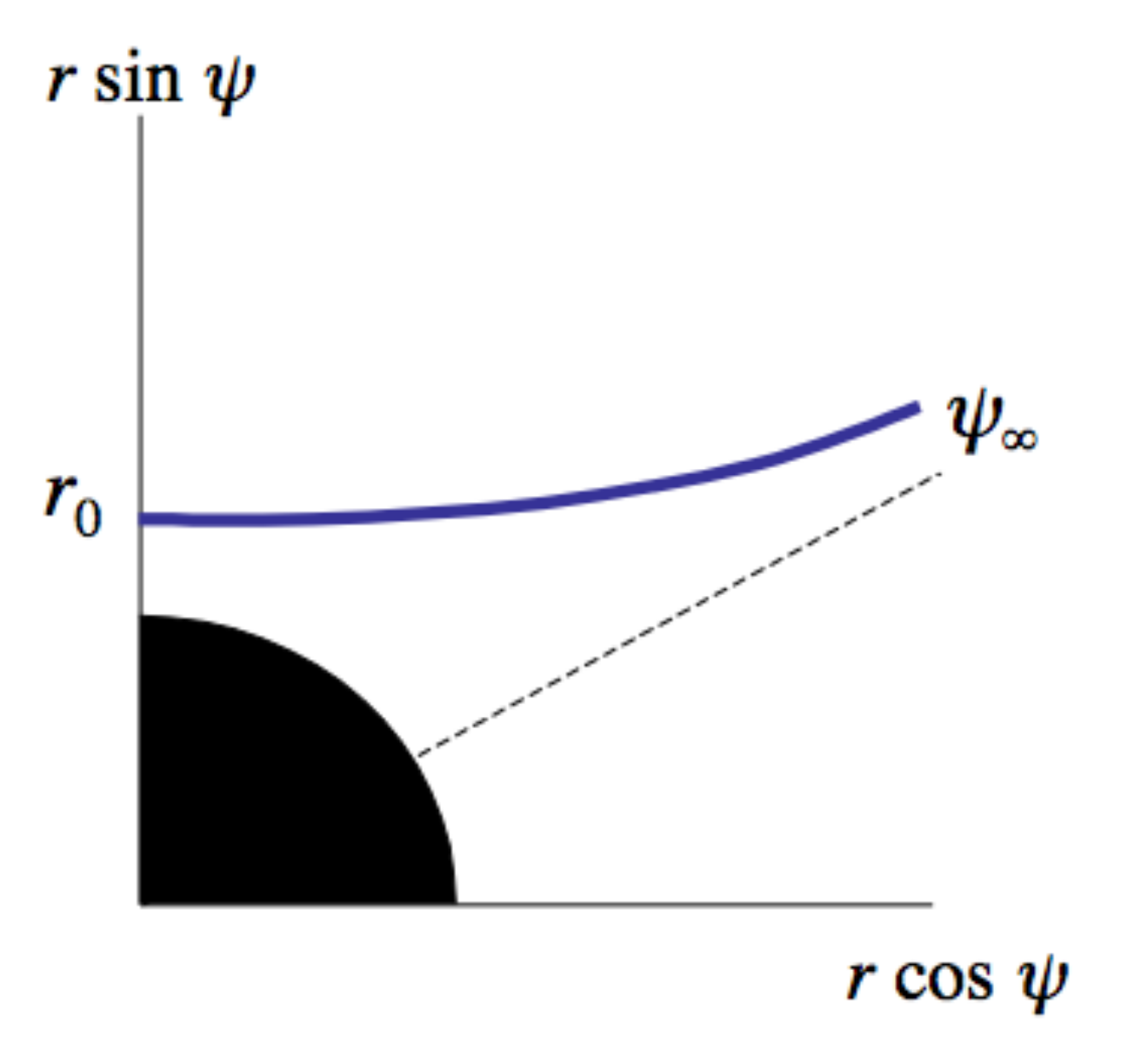,width=4cm} & \hspace{1cm}
\epsfig{file= 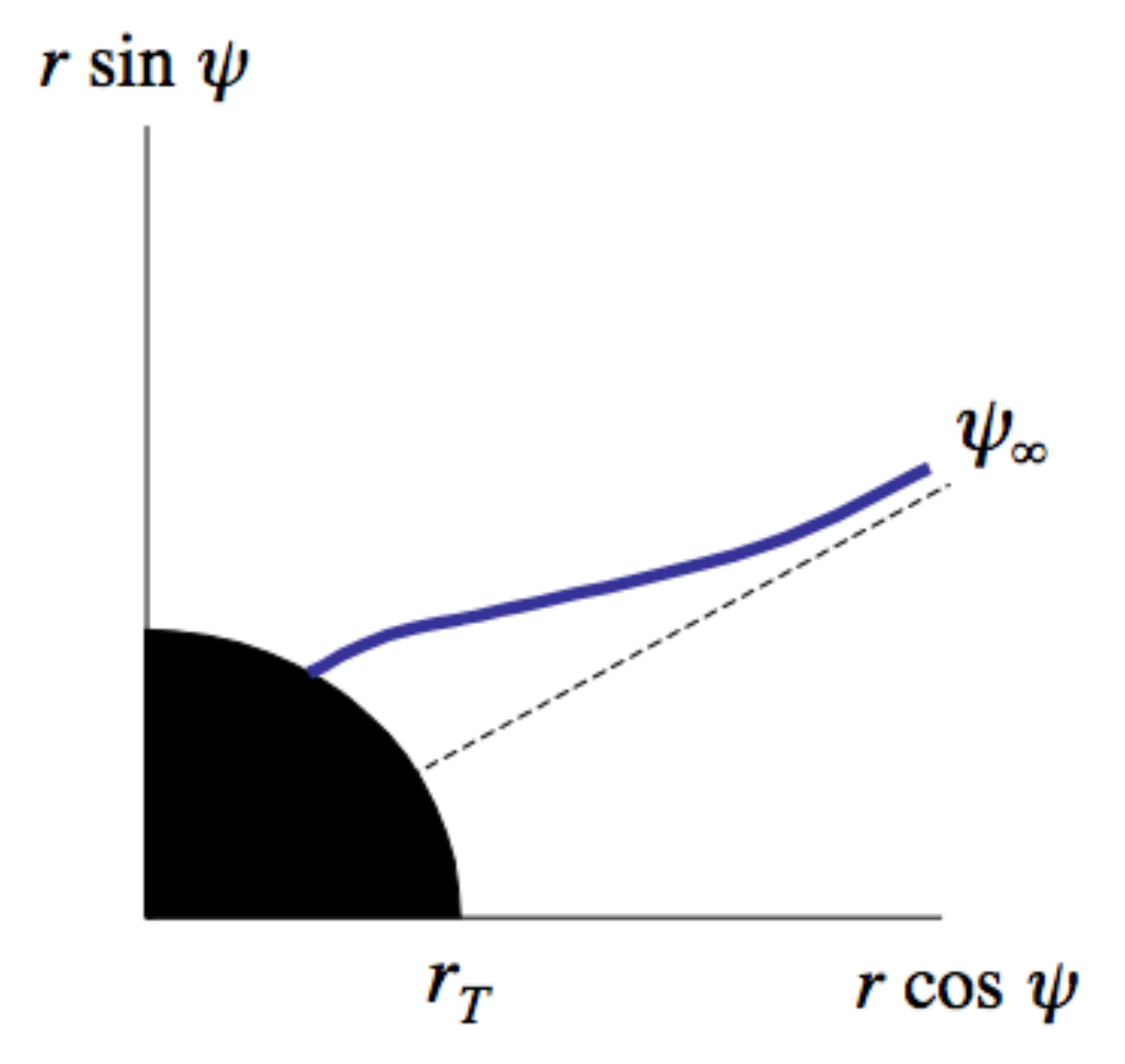,width=4cm}
\end{tabular}
\caption{(a) An MN embedding with $\psi(r_0)=\pi/2$. (b) A BH embedding.}
\label{MN_BH_embeddings}
\end{center}
\end{figure}

Generally there are two types of embbedings, that differ in their small $r$ behavior:
Minkowski-like (MN) embeddings, in which the D7-brane terminates smoothly outside the horizon 
(Fig.~\ref{MN_BH_embeddings}a),
and black-hole (BH) embeddings, in which the D7-brane crosses the horizon (Fig.~\ref{MN_BH_embeddings}b).
We refer the reader to \cite{Bergman:2010gm} for the explicit embedding equations for $\psi(r)$ and $z(r)$,
and for their numerical solutions.

In an MN embedding $\psi(r_0)=\pi/2$ or $0$ for some $r_0>r_T$, corresponding to one or the other $S^2$
shrinking. This indicates that the dual field theory has a mass-gap related to $r_0-r_T$.
An important condition for the existence of MN embeddings is the absence of sources for
the worldvolume gauge field.
Unlike in the model of the previous section, there are no localized sources in this model.
All sources correspond to branes (or strings) connecting the D7-brane to the horizon.
These inevitably pull the D7-brane down to the horizon, resulting in a BH embedding instead.
This means that the flux on the $S^2$ that shrinks must vanish. 
For $\psi(r_0)=\pi/2$, which means that $f_1=0$, 
we find stable massive embeddings in the range 
\be
\label{psi_range}
0.5235 \lesssim \psi_{\infty}  \lesssim 0.6251 \,.
\ee
(There are also the ``mirror" embeddings with $\psi(r_0)=0$ and $f_2=0$.)
Note that the allowed values of $\psi_\infty$ are quantized since the stabilizing flux is quantized.
There are also two isolated MN embeddings with $\psi_\infty=0$ and $\psi_\infty=\pi/2$.

BH embeddings describe gapless phases in the dual theory.
These embeddings exist generically for any $f_1, f_2$ satisfying the stability condition.

\subsection{Finite density and background fields}

For embeddings corresponding to finite density states in a background magnetic field we need to turn on 
the appropriate components of the worldvolume gauge field.
As in the D4-D8 model we will work with the dimensionless field $a_\mu = (2\pi\alpha'/L) A_\mu$.
There are additional terms in the DBI action,
\be
\label{DBI3}
 S_{DBI} & = & - {\cal N} \int dr\, r^2\sqrt{\left(4\cos^4\psi + f_1^2 \right)
 \left(4\sin^4\psi + f_2^2 \right)}\times \nonumber \\ 
 & & \qquad\qquad \times \sqrt{\left(1+ r^4 h(r) z'^2+ r^2 h(r)\psi'^2 - {a_0'}^2\right)\left(1+\frac{b^2}{r^4}\right)} \,,
\ee
and also in the CS action,
\be
\label{CS2}
S_{CS} =  -{\cal N}f_1 f_2 \int dr\, r^4 z'(r)
+ 2{\cal N} \int dr\, c(r) b \, a_0'(r) \,,
\ee
where
\be
\label{c(r)}
c(r) =  \frac{1}{8\pi^2 L^4}\int_{S^2\times S^2} C_4(\psi(r)) 
= \psi(r) - \frac{1}{4}\sin 4\psi(r) -  \psi_\infty + \frac{1}{4}\sin 4\psi_\infty\,.
\ee
We have fixed a gauge for the RR field such that $c(\infty)=0$.
For MN embeddings we must also add a boundary term at $r=r_0$ 
(as explained in section~\ref{D4D8_section})\footnote{In \cite{Bergman:2010gm}
this term was derived by demanding invariance of the CS term $\int C_{4} \wedge F \wedge F$
under gauge transformations of the RR field and then fixing
$c(\infty)=0$. However it can also be obtained by canceling the surface term in the variation of the CS term,
when we present it as $\int F_{5} \wedge A \wedge F$.}
\be
S_{\partial}(r_{0}) =  2{\cal N} c(r_{0}) b\, a_0(r_{0}) \,.
\label{boundary_action_D3D7'}
\ee
The quantity $c(r_{0})$ has a nice physical interpretation: it is the total amount of 5-form flux captured by the 
D7-brane in the MN embedding.
It is completely fixed by the asymptotic value of the embedding angle $\psi_\infty$.
For BH embeddings the boundary term vanishes since $a_0(r_T)=0$.

The integrated equation of motion for $a_0(r)$ is given by
\be
\label{Aeom}
G(r)\, a_0'(r) = d - 2b\, c(r) \,,
\ee
where
\be
\label{G(r)}
G(r) &=& r^2 \left(1 + \frac{b^2}{r^4}\right) \sqrt{\frac{\left(f_1^2+4\cos^4\psi\right)\left(f_2^2+4\sin^4\psi\right)}{Y(r)}}\,, \\
\label{Y(r)}
Y(r) &=& \left(1+\frac{b^2}{r^4}\right) \left(1+hr^4z'^2+hr^2\psi'^2 - a_0^{\prime 2}\right)\,, 
\ee
and $d$ is the total charge density. As in section~\ref{D4D8_section}, we are using (\ref{alternative_current}) to define
the conserved currents.
The quantity on the RHS of (\ref{Aeom}), $\tilde{d}(r)\equiv d - 2bc(r)$, is the contribution to the charge 
density from radial positions below $r$.

We would also like to study the response of the system to a background electric field.
To this end we should consider a more general ansatz for the gauge field with $a_x(t,r)=te + a_x(r)$,
$a_y(x,r)=xb + a_y(r)$, in addition to $a_0(r)$.
The current densities will be contained in the asymptotic behaviors of $a_x(r)$ and $a_y(r)$.
The gauge field equations in this case become
\be
\label{Aeom_2}
G(r)\, a_0'(r) &=& \left[\tilde{d}(r)\left(1 - \frac{e^2}{r^4 h(r)}\right) + \tilde{j}_y(r)\, \frac{eb}{r^4 h(r)}\right]
\frac{1+\frac{b^2}{r^4}}{1+\frac{b^2}{r^4} - \frac{e^2}{r^4 h(r)}} \\
G(r)\, a_y'(r) &=& \left[\tilde{d}(r)\, \frac{eb}{r^4 h(r)} - \frac{\tilde{j}_y(r)}{h(r)}\left(1+\frac{b^2}{r^4}\right)\right]
\frac{1+\frac{b^2}{r^4}}{1+\frac{b^2}{r^4} - \frac{e^2}{r^4 h(r)}} \\
G(r)\, a_x'(r) &=& -\frac{j_x}{h(r)}\left(1+\frac{b^2}{r^4}\right)\,,
\ee
where $j_x$ is the longitudinal current density 
and $\tilde{j}_y(r)$ is defined by analogy with $\tilde{d}(r)$ as $\tilde{j}_y(r) \equiv j_y - 2c(r)e$, 
where $j_y$ is the transverse current density. 
The factor $G(r)$ is defined as before (\ref{G(r)}), now more generally with
\be
Y(r) &=& \left(1+\frac{b^2}{r^4}-\frac{e^2}{hr^4}\right)\left(1+hr^4z'^2+hr^2\psi'^2\right) \nonumber\\
   &   & -\left(1+\frac{b^2}{r^4}\right)a_0'^2+ha_x'^2+\left(1-\frac{e^2}{hr^4}\right)ha_y'^2-\frac{2eb}{r^4}a'_0a'_y \ .
\ee

\subsection{Quantum Hall states}

Let us consider first the response of the gapped MN embeddings.
This is determined by the requirement that there are no sources, namely by regularity of the gauge field at $r=r_0$.
This implies, in particular, that
\be
\label{locking}
\tilde{d}(r_0) = d - 2c(r_0) b = 0 \,.
\ee
The entire charge 
in the MN embedding is thus due to the CS term
and corresponds to a ``fluid" of D5-branes inside the D7-brane.
The charge density is proportional to the magnetic field, and the proportionality constant
is fixed by the value of $c(r_0)$, and therefore of $\psi_\infty$.
This is the key property of a quantum Hall state, which is characterized by a specific 
quantized value of the Landau-level filling fraction $\nu \propto d/b$.
In terms of the physical variables $D=(2\pi\alpha'{\cal N}/L^4)\, d$ and $B=b/(2\pi\alpha')$,
the filling fraction is given by 
\be
\label{filling_fraction}
\nu = {2\pi D\over B} = {2N_c\over \pi} c(r_0)\,.
\ee
For the range of values of $\psi_\infty$ needed for stability (\ref{psi_range}) we get
\be 
 0.6972\lesssim {\nu\over N_c} \lesssim 0.8045 \,.
\ee
Furthermore, the filling fractions are quantized according to the quantization of $\psi_\infty$.
The actual numbers can be obtained by solving (\ref{const_solution}), for a specific flux $f_2$ (with $f_1=0$),
and plugging into (\ref{c(r)}) with $\psi(r_0)=\pi/2$, but they are not particularly illuminating (for example, they are not rational numbers).
The isolated embeddings with $\psi_\infty=0$ and $\psi_\infty=\pi/2$
correspond to $\nu/N_c=1$ and $0$, respectively.

The current densities can likewise be computed by requiring regularity of the spatial components of the gauge field.
This condition implies that
\be
j_x = 0 \;\;\; \mbox{and} \;\;\; \tilde{j}_y(r_0) = j_y - 2c(r_0)e=0 \,,
\ee
from which we can deduce the longitudinal and transverse conductivities:
\be
\sigma_{xx} = 0 \; , \;\; \sigma_{xy} = {\nu\over 2\pi} \,.
\ee
Thus the MN embeddings, when they exist, describe quantum Hall states
with quantized transverse conductivities, and vanishing longitudinal conductivities.
Furthermore, in the holographic description, the quantization is topological since it
originates from the Dirac quantization of the magnetic fluxes on the $S^2$'s.
In particular, $\sigma_{xy}$ in the MN embeddings is independent of the temperature.

\begin{figure}[htbp]
\begin{center}
\begin{tabular}{cc}
\epsfig{file= 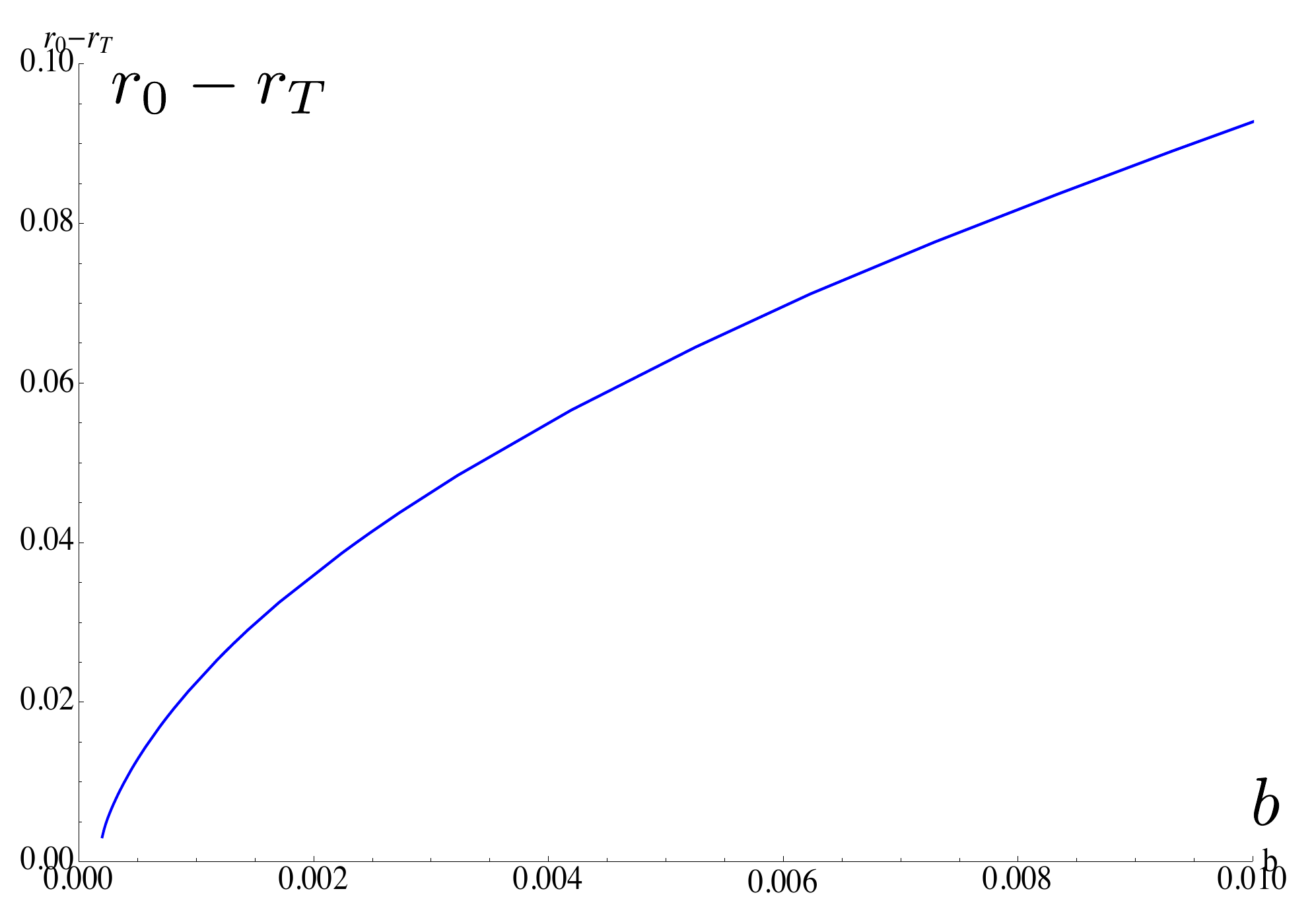,width=5cm} & \hspace{1cm}
\epsfig{file= 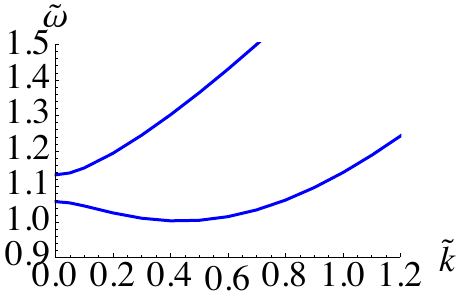,width=5cm}
\end{tabular}
\caption{(a) Mass-gap for charged states as a function of the magnetic field. (b) Dispersion relation of the two lowest neutral modes,
showing the magneto-roton.}
\label{excitations}
\end{center}
\end{figure}

Quantum Hall states are gapped to both charged and neutral excitations.
In this model charged excitations are described by strings 
stretched from the D7-brane to the horizon, and therefore have a mass proportional to $r_0-r_T$.
This is seen to increase with the magnetic field, as shown in Fig.~\ref{excitations}a.
The neutral excitations correspond to fluctuations of the D7-brane worldvolume fields,
and are also found to be massive \cite{Jokela:2010nu} (see also \cite{Jokela:2011sw}). 
The spectrum of neutral excitations includes a magneto-roton, which is a collective excitation
whose dispersion relation has a minimum at non-zero momentum (Fig.~\ref{excitations}b). 
A similar phenomenon is seen in real quantum Hall states \cite{Girvin:1986zz}.

\subsection{Fermi-like liquid}

BH embeddings describe gapless Fermi-like liquids.
For a BH embedding, $\tilde{d}(r_T)$ corresponds to the horizon charge density carried by the ``quarks",
and it need not vanish. In particular, if we add sources to an MN embedding, thereby violating (\ref{locking}), 
it deforms continuously into a BH embedding with horizon charge.

To compute the electrical response in a BH embedding we have a couple of options.
The standard approach is to extract the conductivities using linear response from the current-current correlators,
computed holographically by studying fluctuations of the bulk gauge fields to quadratic order.
The other option is to find a consistent solution in the presence of an external electric field \cite{Karch:2007pd}.
The advantage of the second approach, when it is applicable, is that it gives the complete non-linear response.
Using this method for the BH embeddings one finds
\be
\label{sigma}
\sigma_{xy} & = &
{N_c\over 2\pi^2}\left(\frac{b}{b^2+r_T^4}\tilde{d}(r_T)+2c(r_T)\right) \\
\sigma_{xx} & = & {N_c\over 2\pi^2} {r_T^2\over b^2 + r_T^4}
\sqrt{\tilde{d}(r_T)^2 + (f_1^2+4\cos^4\psi(r_T))(f_2^2+4\sin^4\psi(r_T))(b^2+r_T^4)}\,.
\ee
Note that the transverse conductivity has two components.
The first involves the horizon charge, and resembles the contribution of an ordinary dissipative system of charges.
The remaining charge, corresponding to the fluid of D5-branes inside the D7-brane,
contributes like a dissipationless system.
The longitudinal conductivity involves only the first component.
This is basically the same separation that was seen in the D4-D8 model 
(see equations (\ref{D4D8_long_conductivity}), (\ref{D4D8_trans_conductivity})).

This state of holographic matter exhibits a variety of other interesting phenomena 
as a function of the charge density, temperature, background magnetic field, and mass.

\begin{figure}[ht]
 \center
  \includegraphics[width=0.43\textwidth]{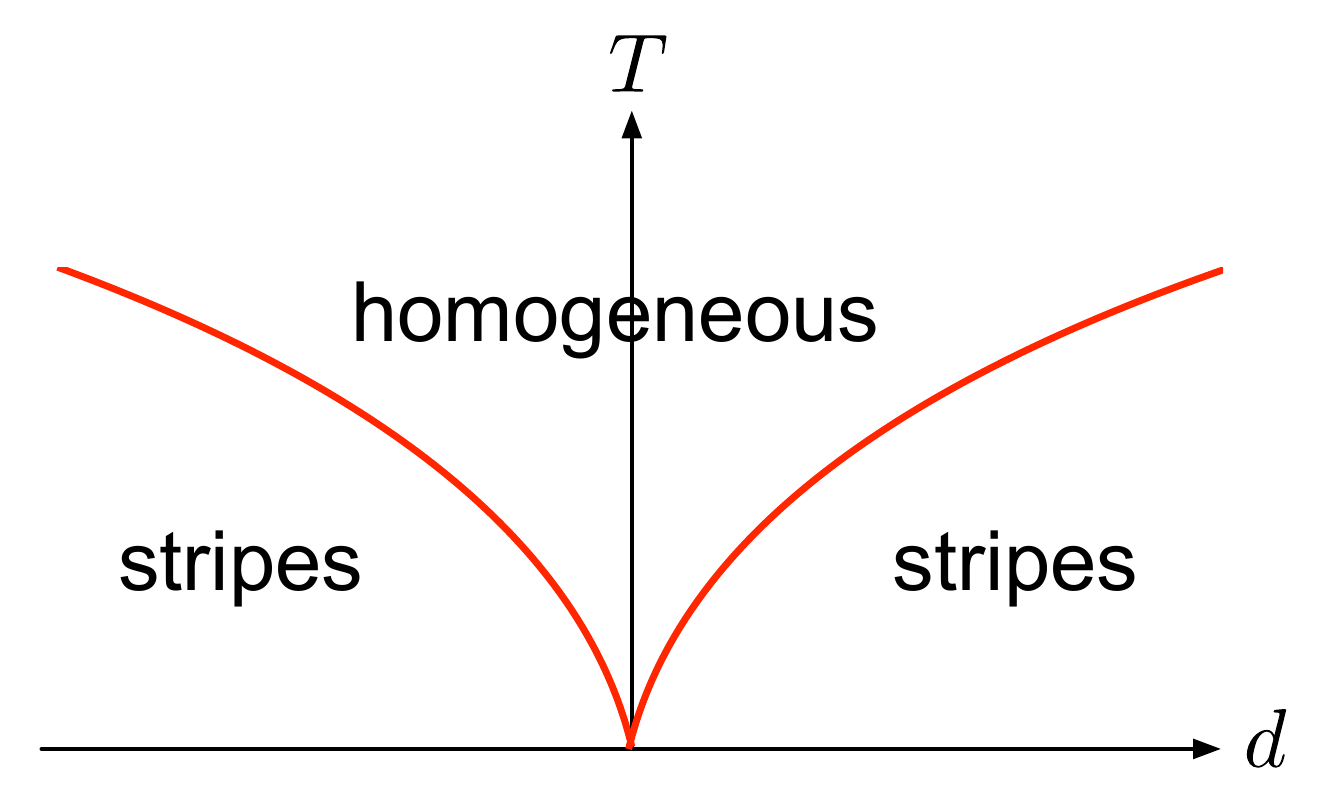}
 \includegraphics[width=0.43\textwidth]{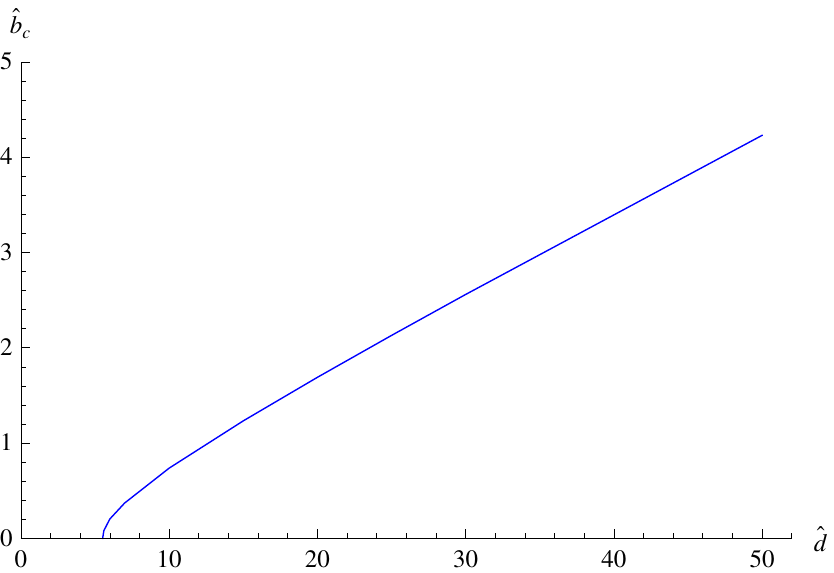}
\caption{(Left)- Phase diagram in the T-d plane showing the quantum critical point. 
(Right)- The phase diagram in the $\hat{b}, \hat{d}$ plane for $m=0$. 
Above the line the system is in the homogeneous phase and below the line in the striped phase.}
\label{magcrit}
\end{figure}

Consider first the state at $T=0$, $d=0$, $b=0$ and $m=0$.
In this case the D7-brane embedding is actually $AdS_{4}\times S^2 \times S^2$, so this situation is 
described by a conformal field theory. Note that in this  case $\sigma_{xx}\neq 0$. 
At non-zero density the system becomes unstable to the formation of stripes. 
The instability  is signaled by the existence of a quasi-normal mode (in the transverse gauge field sector) 
with positive imaginary part in a finite range of momenta \cite{Bergman:2011rf}. 
At a high enough temperature or high enough magnetic field this instability disappears. 
It is convienient to parametrize the situation with $\hat{b}=b/r_{T}^{2}$ and $\hat{d}=d/r_{T}^{2}$. 
Then the instability towards the striped phase at zero magnetic field  and $m=0$ happens for $\hat{d}>5.5$. 
This is demonstrated in Figure~\ref{magcrit}a. 
Above the quantum critical point ($T=m=d=0$) there is a region which resembles a Fermi-like liquid, 
and on both sides there is a striped phase.
For  non-zero $\hat{b}$ and $m=0$ the instability sets in at some other value of $\hat{d}$, as
shown in Fig.~\ref{magcrit}b \cite{Jokela:2012vn} . 
As  $m$ increases the instability sets in at a lower temperature.

\begin{figure}[ht]
 \center
 \includegraphics[width=0.43\textwidth]{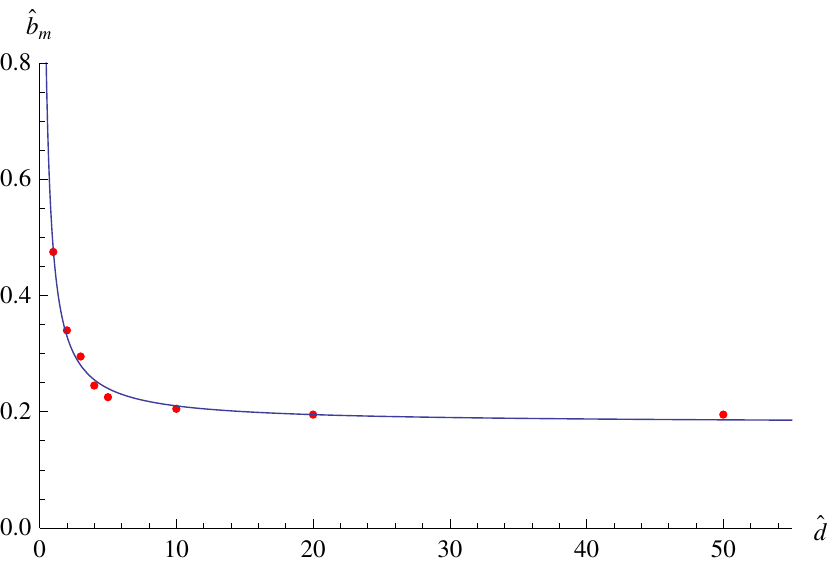}
 \includegraphics[width=0.43\textwidth]{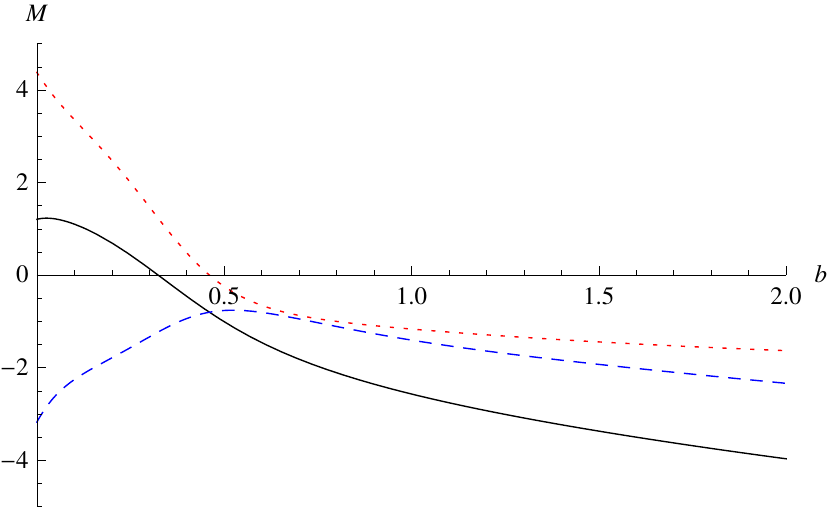}
\caption{(Left)-The line in the $\hat{b}, \hat{d}$ plane  for $m=0$, separating the situation when the 
zero-sound is gapped (above the line) and the situation when it is gapless (below the line). 
(Right)-The magnetization as a function of magnetic field (solid line) and the individual 
contributions from the DBI (dashed line) and the CS term (pointed line) for $m=1$ $d=1$ and $r_{T}=0.1$.}
\label{magneticpro}
\end{figure}

The system also has a zero sound mode. 
At non-zero temperature the quasi-normal mode with the smallest imaginary part at low momentum
is a purely dissipative hydrodynamical mode ($\omega(k=0)=0$).
At some non-zero momentum it
meets another purely dissipative mode ($\omega$ purely imaginary) and crosses from a hydrodynamical 
regime into a collisionless regime, where the resulting complex mode can be identified 
with the finite temperature zero-sound mode \cite{Bergman:2011rf} (see also \cite{Davison:2011ek}). 
The zero sound mode  becomes massive as the magnetic field crosses a critical value 
\cite{Jokela:2012vn, Goykhman:2012vy} (Fig~\ref{magneticpro}a).

For $m \neq 0$ the system can have a non-zero transverse conductivity, even at zero magnetic field. 
This is due to having a non-trivial $c(r)$ for these embeddings. 
This is reminiscent of the anomalous Hall effect (AHE) that appears in ferromagnetic materials
(for a review see \cite{2010RvMP...82.1539N}). 
Indeed for $m \neq 0$ the system is 
ferromagnetic (Fig.~\ref{magneticpro}b) due to the second term in (\ref{CS2}). 
Note that both the AHE and the ferromagnetic behavior, as well as the instability towards a striped phase, 
have a common origin in the Chern-Simon term $\int dr \, c(r) F \wedge F$ in the brane action.

\section*{Acknowledgments}
O.B. and G.L.~would like to thank Matt Lippert and Niko Jokela, who were an integral part
in all our work on the models reviewed in sections 3 and 4. 
O.B. also thanks the Aspen Center for Physics, where this work was completed, for its hospitality.
J.E.~would like to thank her collaborators Martin Ammon, Matthias Kaminski,
Patrick Kerner, Ren\'e Meyer, Jonathan Shock and Migael Strydom,
involved in the joint work presented in section 2.
This work was supported in part by the Israel Science Foundation under grant no.~392/09,
and in part by the US-Israel Binational Science Foundation under grant no.~2008-072.

\bibliographystyle{unsrt}
\bibliography{Magnetic_review_bib_v2.bib}

\end{document}